\documentclass{IEEEtran}

\usepackage{ifpdf}
\usepackage{cite}
\ifCLASSINFOpdf
\usepackage[pdftex]{graphicx}
\DeclareGraphicsExtensions{.pdf,.jpeg,.png}
\else
\usepackage[dvips]{graphicx}
\DeclareGraphicsExtensions{.pdf}
\fi
\usepackage{graphicx}
\ifCLASSOPTIONcompsoc
\usepackage[caption=false, font=normalsize, labelfont=sf, textfont=sf]{subfig}
\else
\usepackage[caption=false, font=footnotesize]{subfig}
\fi
\usepackage{amsthm} % for bold Lemma, Theorem, etc. with italic text
\usepackage{balance}
\usepackage{epstopdf}
\usepackage{amsmath}
\usepackage{amssymb}
\usepackage{color}
\usepackage{balance}
\usepackage{array}
\usepackage{mathtools} % for \splitfrac
\usepackage{algorithm,algorithmic} % for algorithm
\usepackage{enumitem} % for leftmargin
\usepackage[mathcal]{euscript} % follow IEEE proofreading
\usepackage{tabularx} % for table of equations
\usepackage{multirow} % for multiple sub-rows of a cell of a table
\usepackage{booktabs} % for table
\usepackage{makecell} % for thick hline
\usepackage{aligned-overset} % for align overset
\usepackage{url}
\usepackage{multicol} % for two equations side-by-side
\graphicspath{{./Figures/}}
\newtheorem{Remark}{Remark}
\newtheorem{Lemma}{Lemma}
\newtheorem{Theorem}{Theorem}

\renewcommand{\vec}[1]{\boldsymbol{\mathrm{#1}}}

\newcommand{\src}{\mathrm{S}}

\newcommand{\ris}{\mathrm{R}}
\newcommand{\des}{\mathrm{D}}

\newcommand{\idd}{\des}
\newcommand{\ids}{\src}
\newcommand{\idg}{\mathrm{g}}
\newcommand{\idh}{\mathrm{h}}

\newcommand{\idu}{U}

\newcommand{\idx}{X}
\newcommand{\idz}{Z}
\newcommand{\idr}{R}

\newcommand{\era}{\mathrm{ERA}}
\newcommand{\ORA}{\mathrm{ORA}}

\newcommand{\avgSNR}{\bar{\rho}}
\newcommand{\SNRth}{\rho_{\rm th}}

\begin{document}

\title{Multi-RIS-aided Wireless Systems: Statistical Characterization and Performance Analysis}

\author{{Tri~Nhu~Do,~\IEEEmembership{Member,~IEEE,}
		Georges~Kaddoum,~\IEEEmembership{Senior~Member,~IEEE,}
		Thanh~Luan~Nguyen,
		Daniel~Benevides~da~Costa,~\IEEEmembership{Senior~Member,~IEEE,}
		and~Zygmunt~J.~Haas,~\IEEEmembership{Fellow,~IEEE}}%
		\thanks{T.~N.~Do~and~G.~Kaddoum are with the Department of Electrical Engineering, the \'{E}cole de Technologie Sup\'{e}rieure (\'{E}TS), Universit\'{e} du Qu\'{e}bec, Montr\'{e}al, QC H3C 1K3, Canada (emails: tri-nhu.do.1@ens.etsmtl.ca, georges.kaddoum@etsmtl.ca).}
		\thanks{T. L. Nguyen is with the Faculty of Electronics Technology, Industrial University of Ho Chi Minh City, Ho Chi Minh City 700000, Vietnam (e-mail: thanhluan.nguyen.iuh@outlook.com).}
		\thanks{D. B. da Costa is with the Intelligent Wireless Communications (IWiCom) Research Group - Future Technology Research Center, National Yunlin University of Science and Technology, Douliou, Yunlin 64002, Taiwan, R.O.C, and with the Department of Computer Engineering, Federal University of Cear\'{a} (UFC), Sobral 62010-560, Brazil (e-mail: danielbcosta@ieee.org).}
		\thanks{Z. J. Haas is with the Department of Computer Science, University of Texas at Dallas, Richardson, TX 75080, USA, and also with the School of Electrical and Computer Engineering, Cornell University, Ithaca, NY 14853, USA (e-mail: zhaas@cornell.edu).}}

\maketitle

\begin{abstract}
	In this paper, we study the statistical characterization and modeling of distributed multi-reconfigurable intelligent surface (RIS)-aided wireless systems. Specifically, we consider a practical system model where the RISs with different geometric sizes are distributively deployed, and wireless channels associated to different RISs are assumed to be independent but not identically distributed (i.n.i.d.). We propose two purpose-oriented multi-RIS-aided schemes, namely, the exhaustive RIS-aided (ERA) and opportunistic RIS-aided (ORA) schemes. A mathematical framework, which relies on the method of moments, is proposed to statistically characterize the end-to-end (e2e) channels of these schemes. It is shown that either a Gamma distribution or a Log-Normal distribution can be used to approximate the distribution of the magnitude of the e2e channel coefficients in both schemes. With these findings, we evaluate the performance of the two schemes in terms of outage probability (OP) and ergodic capacity (EC), where tight approximate closed-form expressions for the OP and EC are derived. Representative results show that the ERA scheme outperforms the ORA scheme in terms of OP and EC. In addition, under i.n.i.d. fading channels, the reflecting element settings and location settings of RISs have a significant impact on the system performance of both the ERA or ORA schemes.
\end{abstract}


\IEEEpeerreviewmaketitle

\section{Introduction} \label{section_intro}

Reconfigurable intelligent surfaces (RISs) have recently been prospected as one of the key technologies to achieve smart radio environment (SRE) for the sixth generation (6G) of wireless communication systems \cite{DiRenzo_JSAC_2020}. Specifically, in order to create a truly SRE, one of the key considered approaches is to make the radio medium controllable \cite{DiRenzo_JSAC_2020}. 
To this end, RISs, also known as intelligent reflecting surfaces (IRSs) \cite{Wu_TWC_2019} or software-defined metasurfaces (SDMs) \cite{Liaskos_TCOM_2020}, have been developed. In particular, a RIS consists of nearly-passive reflecting elements, which are programmable and controllable via a RIS controller, allowing it to reflect and steer impinging signals toward desired directions. In order to realize this goal, the phase-shifts of the reflecting elements are adjusted, such that the directionality of the beam of scattered signals can be controlled. With this phase-tuning capability, \textit{coherent signal combining} of reflecting signals from different elements of different RISs can be properly implemented such that the multi-path received signals constructively add at the receiver side \cite{Galappaththige_arXiv_2020, Liaskos_TCOM_2020, Tahir_LWC_2020}.

Single-RIS-aided systems have been extensively studied in the literature \cite{Basar_ACCESS_2019,Figueiredo_ACCESS_2021,Badiu_LWC_2020,Boulogeorgos_ACCESS_2020,Gan_LCOMM_2021,Bjornson_LWC_2020_b,VanChienLWC2021}. 
In \cite{Basar_ACCESS_2019}, the authors presented an analytical representation of the ideal phase-shift configuration that maximizes the signal-to-noise ratio (SNR) of the synthesized received signal. 
In \cite{Figueiredo_ACCESS_2021}, considering a large intelligent surface (LIS)-aided single-input single-output (SISO) system without the direct link, the authors showed that the magnitude of the end-to-end (e2e) channel coefficient under Rayleigh fading can approximately follow a Gamma distribution.
In \cite{Badiu_LWC_2020}, similar to \cite{Figueiredo_ACCESS_2021}, considering a large reflecting surface (LRS)-aided SISO system under Rayleigh fading, the authors showed that the magnitude of the e2e channel coefficient approximately follows a Nakagami distribution.
In \cite{Boulogeorgos_ACCESS_2020}, considering a similar system setting as in \cite{Figueiredo_ACCESS_2021} and \cite{Badiu_LWC_2020}, the authors derived approximate closed-form expressions for the probability density function (PDF) and cumulative distribution function (CDF) of the magnitude of the e2e channel coefficient.
In \cite{Gan_LCOMM_2021}, considering a single-RIS-aided system under Rician fading, the authors relied on the central limit theorem (CLT) to show that the magnitude of the e2e channel coefficient (without the direct channel coefficient) approximately follows a Gaussian distribution. 
The aforementioned works made a common assumption that channels associated with different reflecting elements of the same RIS are independent and identically distributed (i.i.d.). 
In \cite{Bjornson_LWC_2020_b}, the authors showed that i.i.d. Rayleigh fading does not physically occur when using a single RIS in an isotropic scattering environment. 
In \cite{VanChienLWC2021, Ibrahim_TVT_2021, Cui_TVT_2021}, relying on the method of moments, the authors showed that the distribution of the e2e channel coefficient of single-RIS-aided systems can be approximated by a Gamma distribution.

In this work, we focus on a more general system setup and consider multi-RIS-aided systems. Next, we discuss some recent papers which assume such a system configuration. 
In \cite{Galappaththige_arXiv_2020}, considering a multi-RIS-aided SISO system with direct link under i.i.d. Nakagami-$m$ fading channels, the authors relied on the CLT to show that the magnitude of the e2e channel coefficient can approximately follow a Gaussian distribution.
In \cite{Jung_TWC_2020}, considering an LIS-aided system under Rician fading, where the LIS consists of RIS units and each RIS unit has a large number of reflecting elements, the authors relied on the law of large numbers and the CLT to approximate the distributions of random variables (RVs), which are functions of the squared magnitude of the channel coefficient, by Gaussian distributions.
In \cite{Yang_WCL_2020}, the authors considered a multi-RIS-aided point-to-point transmission assuming that the direct link does not exist and only the best RIS is selected to participate in the transmission. Assuming that all the Rayleigh fading channels between different RISs are i.i.d., the authors showed that the e2e SNR approximately follows a non-central chi-square distribution.
In \cite{Fang_arXiv_2020}, considering a multi-RIS-aided system, in which the number of elements of each RIS can be arbitrarily adjusted, the authors proposed RIS selection strategies based on the RISs' location information. The performance analysis was carried out based on the assumption that the magnitudes of the channel coefficients associated with different RISs are i.i.d. RVs. 
In \cite{Mei_LWC_2020}, considering multi-hop multi-RIS-aided systems, the authors proposed a RIS selection strategy that maximizes the e2e SNR. However, the authors only took into account the impact of path-loss and ignored the impact of fading.  
In \cite{Zhang_arXiv_2020}, the authors discussed the impact of the centralized and distributed RIS deployment strategies on the capacity region of multi-RIS-aided systems. However, in the distributed RIS deployment, the authors assumed that the channels associated with the different distributed RISs underwent i.i.d. Rayleigh fading. 
In \cite{Yildirim_TCOM_2020}, the authors considered multi-RIS-aided systems for both indoor and outdoor communications, where a direct link between a source and a destination is unavailable; aiming for low-complexity transmission, the authors proposed a RIS selection strategy that selects the RIS with the highest SNR to assist the communication. However, small-scale fading was ignored and the performance analysis of the RIS selection strategy was not carried out. In \cite{Lyu_LWC_2020} and \cite{Lyu_TWC_2021}, the authors evaluated the spatial throughput of single-cell and multi-cell systems, respectively, in which multiple RISs are used to assist multiple users under intra-cell and/or inter-cell interference.

Although previous works have provided important contributions to multi-RIS-aided systems, the accurate characterization of the fading model is still an open problem. Specifically, to facilitate the performance analysis, some existing works only considered path-loss effects and/or ignored small-scale fading effects, as in  \cite{Mei_LWC_2020,Zhang_arXiv_2020,Yildirim_TCOM_2020}. On the other hand, when taking into account small-scale fading, existing works relied on the i.i.d. fading channel model \cite{Galappaththige_arXiv_2020,Yang_WCL_2020}, and \cite{Zhang_arXiv_2020} or deterministic fading channel, as in \cite{Jung_TWC_2020}. 
Channels associated with different elements of the same RIS can be reasonably assumed to be i.i.d. as aforementioned since the elements are typically in sub-wavelength size \cite{DiRenzo_JSAC_2020} and are installed closely to each other on the same panel. However, channels between different RISs cannot be assumed to be i.i.d., because in distributed multi-RIS-aided systems, the RISs are installed significantly far apart, e.g., tens of meters. 

Motivated by the aforementioned observations, in this work, we propose two distributed multi-RIS-aided wireless schemes, namely, the exhaustive RIS-aided (ERA) and opportunistic RIS-aided (ORA) schemes. Specifically, in the former, all the RISs participate in the transmission, whereas in the later, only one scheduled RIS is employed. It is shown that each approach has its own advantages and drawbacks. In particular, the ERA scheme may give better performance at the expense of additional complexity and fronthaul/backhaul load. On the other hand, by using only one RIS, as in the ORA scheme, the other RISs can be employed for other purposes (assisting other pairs of users, for instance), consequently providing a more efficient use of the available resources. Thus, depending on the needs and goals of the application, one approach can be more useful than the other.
Considering such multi-RIS-aided schemes, we raise a research question that still needs to be properly answered in the literature: \textit{what distributions should be used to statistically model the magnitude of the e2e channel coefficient of distributed multi-RIS-aided systems, which is a combination of direct and all reflecting channel coefficients?}
Thus, to answer the raised research question comprehensively, we consider a practical multi-RIS-aided system setting, in which channels associated with reflecting elements of the same RIS are assumed to be i.i.d., whereas channels associated with different RISs are assumed to be independent but not identically distributed (i.n.i.d.), and the system undergoes Nakagami-$m$ fading. 

The key contributions of this work are summarized as follows:
\begin{itemize}

\item The unique technical contribution of our paper lies in our \textit{comprehensive analysis approach}. Specifically, we introduce a framework for implementing the method of moments for statistically characterizing the e2e channel of any wireless system, including distributed multi-RIS-aided wireless systems. Under the considered fading environment, we prove that the distribution of the magnitude of the e2e channel coefficient of the ERA scheme can be approximated by either a Gamma or Log-Normal distribution. On the other hand, the magnitude of the e2e channel coefficient of the ORA scheme can be approximated by a Log-Normal distribution. It is noted that these findings have not yet been reported in the literature. 

\item For the system performance analysis, invoking these findings, tight approximate closed-form expressions for the outage probability (OP) and ergodic capacity (EC) of the ERA and the ORA schemes are derived, based on which insightful discussions are drawn. In addition, we provide an in-depth analysis of the accuracy of the proposed distribution approximation, i.e., the accuracy in using Gamma and Log-Normal distributions to approximate the true distribution of the e2e channel coefficient, based on the lower tail characteristics of the Gamma and Log-Normal distributions, the squared Gaussian RVs-based physical origin, the Kullback-Leibler (KL) divergence, and the Kolmogorov-Smirnov (KS) goodness-of-fit test.

\item In our simulation, we consider realistic simulation settings, as shown in Table~\ref{table_simulation}. Notably, our simulation results are in agreement with the analytical results in all simulation settings. From an engineering (and system design) perspective, our results reveal that the ERA scheme outperforms the ORA scheme in terms of OP and EC; nevertheless, the ORA scheme achieves a higher energy efficiency (EE) in a specific range of the target spectral efficiency (SE) under a certain network setting. From the system infrastructure viewpoint, the number of elements installed at each RIS and the locations of the RISs have a significant impact on the performance of the proposed schemes, in which, the ERA scheme is more robust against changes in the system infrastructure than the ORA scheme. We show that both centralized and distributed multi-RIS-aided systems have their own advantages and drawbacks, which depends on the pre-planed locations of the RISs.
\end{itemize}

{\noindent \bf Notations}:
$\Gamma(\cdot)$ denotes the Gamma function \cite[Eq. (8.310.1)]{Gradshteyn2007}, 
$\gamma(\cdot, \cdot)$ denotes the lower incomplete Gamma function \cite[Eq. (8.350.1)]{Gradshteyn2007},
$K_\nu (z)$ denotes the modified Bessel function of the second kind \cite[Eq. (8.407.1)]{Gradshteyn2007}, $\ln(x) = \log_e (x)$, $\mathrm{erf}(\cdot)$ denotes the error function \cite[Eq. (7.1.1)]{Abramowitz1965}, $G^{m,n}_{p,q} [\cdot]$ denotes the Meijer-G function \cite[Eq. (8.2.1.1)]{Prudnikov1999}, and $\mathrm{erfc}(\cdot)$ denotes the complementary error function \cite[Eq. (8.250.4)]{Gradshteyn2007}. $X \overset{\rm approx.}{\sim} \mathcal{D}(\cdot)$ means $X$ approximately follows the distribution $\mathcal{D}(\cdot)$. For parameters of distributions, e.g., ${\alpha_X}$ means the parameter $\alpha$ in the distribution of random variable $X$, $\mu_X (k)$ denotes the $k$-th moment of $X$, and $F_A (\cdot)$ denotes the Lauricell function Type-A \cite[Eq.(1.4.1)]{Srivastava1985}. $\mathbb{E}[X]$ and $\mu_X$ denote the expectation of $X$ and the corresponding expected value.

\section{System Model} \label{section_system}

\begin{figure} [t]
\centering
\includegraphics[width=.9\linewidth]{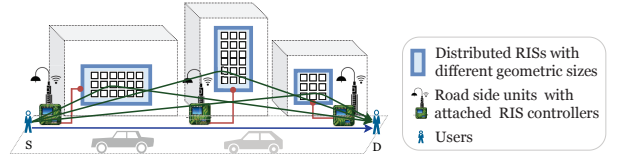}
\caption{Illustrations of a distributed multi-RIS-aided wireless system, where RISs can be installed on houses and/or buildings.}
\label{fig1_system}
\end{figure}

We consider a distributed multi-RIS-aided system, consisting of one source, $\src$, that communicates with one destination, $\des$, through a direct link with the assistance of $N$ distributed RISs, $\ris_n, n=1,\ldots,N$, as depicted in Fig. \ref{fig1_system}. It is assumed that $\src$ and $\des$ are single-antenna nodes, whereas the $n$-th RIS is equipped with $L_n$ passive reflecting elements. It is noted that the RISs may have different geometric sizes, i.e., $L_i \neq L_j, i,j \in \{1,\ldots,N\}$. Let $\vec{\Theta}_n = \mathrm{diag}([\kappa_{n1} e^{j \theta_{n1}}, \ldots, \kappa_{nl} e^{j \theta_{nl}}, \ldots, \kappa_{nL_n} e^{j \theta_{nL_n}}])$ denote the phase-shift matrix \cite{Bjornson_WCL_2020}, where $\kappa_{nl} \in (0,1]$ and $\theta_{nl} \in [0, 2\pi)$ stand for, respectively, the amplitude reflection coefficient and the phase-shift of the $l$-th reflecting element of the $n$-th RIS. Each RIS is programmed by a controller, where the controllers can be connected, for example, via a high-speed backhaul to synchronize the phase-shift configuration over the entire system. 
To provide a more general system model and comprehensive performance analysis, the direct link from $\src$ to $\des$ is assumed to be available. Accordingly, the use of RISs aims to increase the reliability of the direct transmission by creating additional propagation paths.
Furthermore, we assume that perfect global channel state information (CSI) is available at the RIS controllers and the source.
In order to obtain the CSI, we resort to channel estimation techniques proposed for (multi)-RIS-aided systems in the literature, such as in \cite{Zheng_LWC_2020,AlwazaniGLOBECOM2020}.\label{page_channel_estimation}
It is noted that a possible deployment of the distributed multi-RIS-aided systems is over quasi-static wireless channels in low mobility environments, such as walking pedestrian environments.
In addition, we assume interference-free signal reflection between RISs. For instance, one possible scenario is that RISs are deployed as uniform rectangular arrays (URAs) implanted on a wall, which leads to no interference between the RISs.

Let $\tilde{h}_{nl}$ and $\tilde{g}_{nl}$ denote the \textit{complex channel coefficients} from $\src$ to the $l$-th reflecting element of the $n$-th RIS, and from that reflecting element to $\des$, respectively; and let $\tilde{h}_0$ denote channel coefficient of the direct $\src \to \des$ link. 
Knowing that, for a complex number $z$, one can represent $z = r (\cos\theta + j \sin\theta) = r e^{j\theta}$, where $\theta = \angle z$ and $r = |z|$, and $|r e^{j\theta}| = |r e^{-j\theta}|, \forall r \in \mathbb{R}$, the channels can be expressed in their polar representations as $\tilde{h}_0 = h_0 e^{j \phi_0}$, $\tilde{h}_{nl} = h_{nl} e^{j \phi_{nl}}$, $\tilde{g}_{nl} = g_{nl} e^{j \psi_{nl}}$, where $h_0$, $h_{nl}$, and $g_{nl}$ are \textit{the magnitudes of the channel coefficients}, i.e., $h_0 = |\tilde{h}_0|$, $h_{nl} = |\tilde{h}_{nl}|$, and $g_{nl} = |\tilde{g}_{nl}|$, and $\phi_0$, $\phi_{nl}$, and $\psi_{nl}$ are \textit{the phases} of $\tilde{h}_0$, $\tilde{h}_{nl}$, and $\tilde{g}_{nl}$, respectively, with $\{\phi_0, \phi_{nl}, \psi_{nl}\} \in [0,2\pi]$.
Let $x_\ids$, with $\mathbb{E}[|x_\ids|^2] = 1$, be the transmit symbol at $\src$, $P_\ids$ be the transmit power (in [dBm]) of $\src$, and $w_\idd$ be the additive white Gaussian noise (AWGN) at $\des$ with zero mean and variance $\sigma_\idd^2$, i.e., $w_\idd \sim \mathcal{CN} (0, \sigma_\idd^2)$.

\subsection{The Exhaustive RIS-aided Scheme} \label{subsec_ERA_scheme}

In the ERA scheme, all the RISs assist the transmission between $\src$ and $\des$, i.e., when $\src$ transmits its signal $x_\ids$ to $\des$, all $N$ RISs are controlled to reflect replicas of $x_\ids$ to $\des$ over the same time-frequency channel. 
Herein, we assume that all RISs are controlled to steer their beams to the destination, and therefore they cannot interfere with each other, as in \cite{Galappaththige_arXiv_2020,Yang_WCL_2020,Fang_arXiv_2020}, and  \cite{Yildirim_TCOM_2020}. As a result, $D$ receives a superposition (combination) of the all received multi-path signals. Thus, using coherent combination \cite{Galappaththige_arXiv_2020} and \cite{VanChienLWC2021}, the received signal at $\des$, which is synthesized from the direct signal and reflected signals from all $N$ RISs, can be written as
\begin{align} \label{eq_rx_signal}
y^\era = \sqrt{P_\ids} \bigg( \tilde{h}_0 + \sum_{n=1}^N \sum_{l=1}^{L_n} \tilde{g}_{nl} \kappa_{nl} e^{j \theta_{nl}} \tilde{h}_{nl} \bigg) x_\ids + w_\idd.
\end{align}

Using polar representation of the complex channel coefficients, the received SNR at $\des$ can be expressed as
\begin{align} 
    \mathrm{SNR}^\era 
    &= \avgSNR \bigg\vert h_0 e^{j \phi_0} + \sum_{n=1}^N \sum_{l=1}^{L_n} g_{nl} \kappa_{nl} h_{nl} e^{j (\theta_{nl} + \phi_{nl} + \psi_{nl})} \bigg\vert^2 \nonumber \\
    &= \avgSNR \underbrace{|e^{ j \phi_0}|^2}_{=1} \bigg\vert h_0 + \sum_{n=1}^N \sum_{l=1}^{L_n} g_{nl} \kappa_{nl} h_{nl} e^{j \delta_{nl}} \bigg\vert^2,
\end{align}
where $\avgSNR = P_\ids/\sigma_\idd^2$ denotes the \textit{average transmit SNR} [dB] and $\delta_{nl} \triangleq \theta_{nl} + \phi_{nl} + \psi_{nl} - \phi_0$ is \textit{the phase error} of the $l$-th reflecting element of the $n$-th RIS.

In order to overcome the destructive effect of multi-path fading, the phase-shifts of the RISs are reconfigured such that all the received signals are constructively added to achieve the largest SNR. Mathematically, the ideal phase-shift configuration of the $l$-th reflecting element of the $n$-th RIS can be expressed as 
\begin{align}
 \theta_{nl}^* = \arg\max_{\theta_{nl} \in \mathcal{Q}_n} \mathrm{SNR}^\era (\theta_{nl}), \forall l,\forall n.
\end{align}
In practice, considering discrete phase-shifts, the number of phase-shifts is limited and constrained by a phase-shift resolution, which is the number of quantization bits, $b_n$, for the $n$-th RIS. Let $Q_n \triangleq 2^{b_n}$ denote the \textit{phase-shift resolution} of the $n$-th RIS. Thus, the value of a phase-shift can be picked from the following set $\mathcal{Q}_n = \left\{0, \frac{2 \pi}{Q_n}, \frac{4 \pi}{Q_n}, \ldots, \frac{2 \pi (Q_n - 1)}{Q_n}\right\}$ \cite{Huang_TWC_2019}. 
Assuming a high phase-shift resolution, i.e., $2^{b_n} \gg 1$, and perfect channel state information (CSI), the $n$-th RIS is able to generate ideal phase-shifts such that the phase errors can be zero, i.e., $\delta_{nl} = 0, \forall l, \forall n$. It is noted that the ideal phase-shift assumption has been extensively used for single-RIS-aided systems \cite{Basar_ACCESS_2019, Boulogeorgos_ACCESS_2020, Yang_TVT_2020, Gan_LCOMM_2021}, and \cite{Bjornson_LWC_2020_b}, multi-RIS-aided systems \cite{Galappaththige_arXiv_2020} and \cite{Yang_WCL_2020}, and multi-hop multi-RIS-aided networks \cite{Mei_LWC_2020}.
As a result,  $\theta_{nl}^* =  \phi_0 - (\phi_{nl} + \psi_{nl}) $, which yields a synthesized received signal at $\des$ having the largest amplitude. Note that since $ e^{j (\theta + 2k \pi)} = e^{j \theta}$, $\theta_{nl}^* \in [0, 2\pi)$. Thus, the received SNR at $\des$ can be re-expressed as
\begin{align} \label{SNR_ERA}
\mathrm{SNR}^\era = \avgSNR \bigg\vert h_0 + \sum_{n=1}^{N} \sum_{l=1}^{L_n} \kappa_{nl} h_{nl} g_{nl} \bigg\vert^2.
\end{align} 

\subsection{The Opportunistic RIS-aided Scheme}

Instead of using all the RISs, we propose the ORA scheme with the aim to reduce the resource usage. In this case, only the most appropriate RIS participates in assisting the direct transmission. The benefit of the ORA scheme is that it provides a low-complexity and energy-efficient transmission protocol at the receiver side. Consequently, the receiver in the ORA scheme handles a significantly lower number of reflecting signals compared to the ERA scheme. Specifically, assuming that the $n$-th RIS is scheduled to assist the direct transmission, the received signal at $\des$ can be written as
\begin{align}
y^\ORA_n = \sqrt{P_\ids} \bigg( \tilde{h}_0 + \sum_{l=1}^{L_n} \tilde{g}_{nl} \kappa_{nl} e^{j \theta_{nl}} \tilde{h}_{nl} \bigg) x_\ids + w_\idd .
\end{align}
The received SNR associated with the $n$-th RIS can be expressed as
\begin{align}
 \mathrm{SNR}^\ORA_n = \avgSNR \underbrace{|e^{j \phi_0}|^2}_{=1} \bigg\vert h_0 + \sum_{l=1}^{L_n} g_{nl} \kappa_{nl} h_{nl} e^{j \delta_{nl}} \bigg\vert^2,
\end{align}
where $\delta_{nl} = \theta_{nl} + \phi_{nl} + \psi_{nl} - \phi_0$. In the ORA scheme, the selected RIS is the one that potentially provides the highest e2e received SNR at $\des$. Specifically, the opportunistic scheduling criterion to choose the most appropriate RIS can be mathematically expressed as
\begin{align} \label{selection_criterion_a}
	n^* = \arg\max_{1\leq n \leq N} \max_{ \theta_{nl} \in \mathcal{Q}_n } \mathrm{SNR}^\ORA_n,  \forall l, \forall n .
\end{align}
In order to meet \eqref{selection_criterion_a}, the ideal phase-shift configuration is first estimated at each reflecting element of each RIS, i.e., $\theta_{nl}^* = \phi_0 - (\phi_{nl} + \psi_{nl}), \forall l, \forall n $. Thus, the criterion \eqref{selection_criterion_a} can be rewritten as $	n^* = \arg\max_{1\leq n \leq N} \mathrm{SNR}^\ORA_n$, i.e.,
\begin{align} \label{SNR_ORA_n}
\mathrm{SNR}^\ORA_{n^*} 
&=  \max_{1 \leq n \leq N} \mathrm{SNR}^\ORA_n \nonumber \\
&= \max_{1 \leq n \leq N} \bigg\{  \avgSNR \bigg\vert h_0 + \sum_{l=1}^{L_n} g_{nl} \kappa_{nl} h_{nl} \bigg\vert^2  \bigg \}.
\end{align}

As can be observed, the operation of the ORA scheme requires the same number of CSI estimations as required by the ERA scheme. Nevertheless, the destination node in the ORA scheme just needs to process $(L_n + 1)$ received signals compared to $(N \times L_n + 1)$ signals in the ERA scheme. This difference in the participating passive elements results in a significant gap between the EE of the two schemes.

\section{Performance Analysis of the ERA Scheme}\label{section_analysis_ERA}

For notational simplicity, along the analysis, let 
$U_{nl} \triangleq \kappa_{nl} h_{nl}g_{nl}$,
$V_n \triangleq  \sum_{l=1}^{L_n} U_{nl}$,
$M_V \triangleq \max_{1 \leq n \leq N} V_n$,
$T \triangleq \sum_{n=1}^{N} V_n$,
$Z \triangleq h_0 + T$,
and $R \triangleq h_0 + M_V$.
The magnitudes of the e2e channel coefficients in the ERA and the ORA schemes can be represented by the RVs $Z$ and $R$, respectively. Directly deriving true distributions of these RVs is infeasible because of their complicated structures. To deal with this technical problem, we propose a methodology to determine an approximate version of the true distribution.

\subsection{The Proposed Framework for Statistical Characterization of the Magnitude of the e2e Channel}

The 3-step framework is proposed as follows:
 
\begin{itemize}
\item Step 1: We first determine which parametric distribution is the best candidate to approximate the true distribution. Specifically, we heuristically and numerically match the \textit{simulated true distribution} to some known candidate distributions, e.g., Gamma distribution, Log-Normal distribution, Gaussian distribution, Burr distribution, Weibull distribution, just to name of few, by using the $\mathrm{fitdist}$ function of Matlab \cite{Matlab2021a}. We then select the best candidate distributions, which give a high goodness of fit and accurate simulation results of performance metrics of interests, e.g., outage probability (OP) and/or ergodic capacity (EC), as will be shown in Theorems \ref{theorem_dist_Z_Gamma}, \ref{theorem_dist_Z_LogNormal}, and \ref{theorem_dist_R_LogNormal}. If no appropriate candidate distribution can be found, we focus on matching the simulated true distribution of a key component of the considered RV to the known candidate distribution, as will be shown in Theorem \ref{theorem_dist_R_Gamma}. 
\item Step 2: We next determine the statistical characteristics, i.e., parameters, of the candidate parametric distribution using \textit{the method of moments} \cite{Bowman1998}. In particular, based on the knowledge of the statistical characteristics of the true distribution of individual components, e.g., the $k$-th moment of the Nakagami-$m$ distribution of each individual channel, we determine the probability distribution and the associated parameters of the candidate distribution that best fits the statistical characteristics of the magnitude of the e2e channel coefficient. The key derivation is to determine \textit{the method of moments estimators} of the parameters of the candidate distribution. More specifically, to derive the estimators, we match the first $k$ moments of the candidate distribution, which are unknown, with that of the simulated true distribution, which are known as \textit{population moments} and can be derived, to form a system of equations that can be solved to find the representation of the estimators. For instance, if the candidate distribution of the magnitude of the e2e channel coefficient is postulated to be a parametric Gamma distribution with the PDF given in \eqref{PDF_Gamma}, the tricky part is how to determine the estimators for the parameters $\alpha$ and $\beta$ of this Gamma distribution, e.g., solve the system of equations and find the $k$-th moment of the simulated true distribution, such that this distribution can accurately characterize the true distribution of the magnitude of the e2e channel coefficient. 
\item Step 3: We verify the accuracy of the obtained approximate distribution. Specifically, in order to evaluate the accuracy of the obtained statistical model, we rely on the corroboration between the simulated true distribution and the matched distribution.
\end{itemize}

We elaborate more on the advantages of the proposed framework as follows. 
According to the definition of i.i.d. RVs, as in \cite[pp. 239]{Peebles2000}, the i.i.d. RVs have the same distribution with the same statistical characteristics. For instance, let $X_1$ and $X_2$ are Nakagami-$m$ independent RVs, i.e., $X_1 \sim \mathrm{Nakagami} (m_1, \Omega_1)$ and $X_2 \sim \mathrm{Nakagami} (m_2, \Omega_2)$. If $X_1$ and $X_2$ are i.i.d. RVs, then $m_1 = m_2 = m$ and $\Omega_1 = \Omega_2 = \Omega$. If $X_1$ and $X_2$ are i.n.i.d. RVs, then there is no restriction between $m_1$ and $m_2$, meaning that $m_1$ and $m_2$ can be identical ($m_1 = m_2$) or non-identical ($m_1 \ne m_2$). This relation also applies to $\Omega_1$ and $\Omega_2$. Directly applying the method of moments to either $Z$ or $R$ is infeasible because of their complicated structures. For instance, in the ERA scheme, $Z$ is the sum of $N$ i.n.i.d. RVs, where each of these RVs consists of $2 L_n$ i.i.d. RVs. Our proposal framework is more flexible in the sense that it can apply the method of moments to any small parts of $Z$ or $R$, and gradually leads to an ultimate approximate distribution of $Z$ or $R$.

It is noteworthy that, for large-scale systems with co-channel inter-cell interference, our proposed framework can be used as follows. Instead of dealing directly with the e2e signal-to-interference-plus-noise ratio (SINR), we deal with individual RVs in the numerator and denominator of the SINR expression. Specifically, we use the framework to find the approximate distributions of the true distributions of these individual RVs. The method of moments can be iteratively applied to the approximate distributions, leading to the ultimate approximate distribution, which facilitates the performance analysis of interest. It is noted that the more distribution fittings are performed, the more inaccurate the approximate distribution is. Nevertheless, this inaccuracy can be compensated by increasing the degree of freedom of the e2e channel, e.g., by installing more reflecting elements.

Next, we present some distributions, which will be used along the performance analysis.
Let $X$ be a RV following a Nakagami-$m$ distribution with its PDF and CDF, parameterized by $m$ and $\Omega$, given by \cite{Peebles2000}
\begin{align}
	f_X(x; m, \Omega) &= \frac{2{m}^{m}}{\Gamma(m){\Omega}^{m}} x^{2m -1} e^{-\frac{m}{\Omega} x^2},  \label{PDF_Naka} \\
	F_X (x;m , \Omega) &= \frac{\gamma \left(m, \frac{m}{\Omega} x^2 \right)}{ \Gamma(m)}. \label{CDF_Naka}
\end{align}
Here, $m > 0$ is the \textit{shape parameter}, indicating the severity of fading and $\Omega > 0$ is the \textit{spread parameter} of the distribution. Next, we use the alternative representation for denoting a Nakagami-$m$ RV: $ X \sim \mathrm{Nakagami}(m, \Omega)$.
It is noted that $\Omega$ denotes the mean square value of $X$, i.e., $\Omega = \mathbb{E}[X^2]$ \cite{Bithas_TCOM_2020}, which is equivalent to the average channel (power) gain. 
The distribution of the magnitude of each individual channel can be expressed as $h_0 \sim \mathrm{Nakagami}(m_0, \Omega_0)$, $h_{nl} \sim \mathrm{Nakagami}(m_{\idh_n}, \Omega_{\idh_n})$, and  $g_{nl} \sim \mathrm{Nakagami}(m_{\idg_n}, \Omega_{\idg_n})$, where $l = 1,\ldots,L_n$ and $n = 1,\ldots,N$.

Let $Y$ be a RV following a Gamma distribution, whose PDF and CDF, parameterized by $\alpha$ and $\beta$, respectively given by \cite{Peebles2000}
\begin{align}
	f_Y (y ; \alpha, \beta) &= \frac{{\beta}^{\alpha}}{\Gamma(\alpha)} y^{\alpha - 1} e^{-\beta y}, y \geq 0, \label{PDF_Gamma} \\
	F_Y (y ; \alpha, \beta) &= \frac{ \gamma (\alpha, \beta y) }{ \Gamma(\alpha) }, y \geq 0 . \label{CDF_Gamma}
\end{align}
Here, $\alpha > 0$ is the \textit{shape parameter} and $\beta > 0$ is the \textit{rate parameter} of the distribution. Hereafter, we use the following representation to denote a Gamma RV: $Y \sim \mathrm{Gamma}(\alpha,\beta)$.

Let $W$ be a RV following a Log-Normal distribution, whose PDF and CDF are given by \cite{Peebles2000}
\begin{align}
	f_{W} \left( w; \nu, \zeta \right) 
	&= \frac{1}{ w \sqrt{2 \pi \zeta^2} } 
	e^{ -\frac{ \left( \ln w - \nu \right)^2 }{ 2 \zeta^2 } }, \label{PDF_log_normal} \\
	F_{W} \left( w; \nu, \zeta \right) 
	&= \frac{1}{2} 
	+ \frac{1}{2} \mathrm{erf} 
	\left( \frac{ \ln w - \nu }{ \sqrt{2 \zeta^2} } \right), 
	\label{CDF_log_normal}
\end{align}
respectively, where $\nu$ and $\zeta^2$, with $\zeta > 0$, are the \textit{mean} and the \textit{variance} of the distribution of $W$. Hereafter, we use the following representation to denote a Log-Normal RV: $W \sim \mathrm{LogNormal}(\nu, \zeta)$.

\subsection{Statistical Channel Characterization of the ERA Scheme Based on Gamma Distribution}\label{subsection_ERA_Gamma}

Relying on the proposed distribution estimation framework, we show that the true distribution of $Z$ can be accurately approximated by the Gamma distribution, as shown in Theorem \ref{theorem_dist_Z_Gamma}. 

\begin{Theorem} \label{theorem_dist_Z_Gamma}
The true distribution of $Z$ can be approximated by the Gamma distribution, which is characterized by two parameters $\alpha_\idz$ and $\beta_\idz$, i.e., $Z \overset{\rm approx.}{\sim} \mathrm{Gamma} ({\alpha_\idz}, {\beta_\idz})$, where the estimators of $\alpha_\idz$ and $\beta_\idz$ can be expressed as
\begin{align}
	{\alpha_\idz} &= \frac{(\mathbb{E}[Z])^2}{\mathrm{Var}[Z]} = \frac{[\mu_Z(1)]^2}{\mu_Z(2) - [\mu_Z(1)]^2}, \label{alpha_Z} \\
	{\beta_\idz} &= \frac{\mathbb{E}[Z]}{\mathrm{Var}[Z]} = \frac{\mu_Z(1)}{\mu_Z(2) - [\mu_Z(1)]^2}, \label{beta_Z}
\end{align}
respectively, where $\mu_Z (1)$ and $\mu_Z (2)$ are presented in \eqref{mu_Z_1} and \eqref{mu_Z_2}, respectively. Thus, the approximate PDF and CDF of $Z$, i.e., $f_{Z} (z; {\alpha_\idz}, {\beta_\idz})$ and $F_{Z} (z; {\alpha_\idz}, {\beta_\idz})$, can be expressed using \eqref{PDF_Gamma} and \eqref{CDF_Gamma}, respectively.
\end{Theorem}

\begin{IEEEproof}
The proof is provided in Appendix \ref{proof_theorem_dist_Z_Gamma}.
\end{IEEEproof}

It is worth noting that the tricky part is to determine the statistical characteristics of the Gamma distribution, i.e., $\mu_Z (1)$ and $\mu_Z (2)$. In addition, knowing that for arbitrary $X$ and $Y$, where $Y = c X^2$, we have $F_Y(y) = F_X (\sqrt{y/c})$ and $f_Y (y) = \frac{1}{2 \sqrt{cy}} f_X \left( \sqrt{\frac{y}{c}} \right)$, the PDF and CDF of $Z^2$ can be obtained as 
\begin{align}
	f_{Z^2} (x) &\approx 
	\frac{1}{2 \sqrt{x}} \frac{{\beta_\idz}^{\alpha_\idz}}{\Gamma(\alpha_\idz)} \! \left( \sqrt{x} \right)^{\alpha_\idz - 1} \! \! e^{-\beta_\idz \sqrt{x} }, \label{CDF_Z2_c} \\
	F_{Z^2} (x) &\approx 
	\frac{ \gamma \left( \alpha_\idz, \beta_\idz \sqrt{x} \right) }{ \Gamma(\alpha_\idz) }, \label{CDF_Z2_e}
\end{align}
respectively. On the other hand, the PDF and CDF of the Generalized Gamma (GG) distribution are, respectively, given by \cite{Stacy1962}
\begin{align}
	f_{\mathrm{GG}}(x;a,d,p) &=  \frac{ \left( p / a^d \right) x^{d-1} e^{- \left( \frac{x}{a}\right)^p} }{ \Gamma \left( d/p\right) }, \label{PDF_generalized_gamma} \\
	F_{\mathrm{GG}}(x; a, d, p) &= \frac{ \gamma \left( d/p, \left( x/a \right)^p\right)}{\Gamma(d/p)}, \label{CDF_generalized_gamma}
\end{align}
where $a, d, p$ are parameters of the distribution. Thus, \eqref{CDF_Z2_c} can be rewritten using \eqref{PDF_generalized_gamma} as
\begin{align} 
	f_{Z^2} (x) &= \frac{[1/2] / \left[ \left(\beta_\idz^{-2}\right)^{\frac{\alpha_\idz}{2}} \right]}{ \Gamma( [\alpha_\idz/2]/[1/2] )} 
	x^{\frac{\alpha_\idz}{2} - 1} 
	e^{- \left( \frac{x}{\beta_\idz^{-2}} \right)^{\frac{1}{2}} } \label{CDF_W_d}.
\end{align}

Next, we present the statistical characterization of the \textit{e2e channel power gain} of the ERA scheme. By comparing \eqref{PDF_generalized_gamma} and \eqref{CDF_W_d}, the distribution functions of $Z^2$ can be represented in the form of the Generalized Gamma distribution as in the following Remark \ref{remark_Generalized_Gamma}.

\begin{Remark} \label{remark_Generalized_Gamma}
From Theorem \ref{theorem_dist_Z_Gamma}, the true distribution of $Z^2$ can be approximated by using the Generalized Gamma distribution, where the PDF and CDF of $Z^2$ can be expressed as
\begin{align}
	f_{Z^2} (x) &= f_{\mathrm{GG}}\left(x ; a_{\idz^2},d_{\idz^2},p_{\idz^2} \right), \label{PDF_Z2_GG_a} \\
	F_{Z^2} (x) &= F_{\mathrm{GG}} \left(x; a_{\idz^2},d_{\idz^2},p_{\idz^2} \right), \label{CDF_Z2_GG_a}
\end{align}
respectively, where $f_{\mathrm{GG}}(\cdot)$ and $F_{\mathrm{GG}}(\cdot)$ are presented in \eqref{PDF_generalized_gamma} and \eqref{CDF_generalized_gamma}, respectively, $a_{\idz^2} = \beta_\idz^{-2}$, $d_{\idz^2} = {\alpha_\idz}/2$, and $p_{\idz^2} = 1/2$, where ${\alpha_\idz}$ and ${\beta_\idz}$ are expressed in \eqref{alpha_Z} and \eqref{beta_Z}, respectively.
\end{Remark}

\subsubsection{Outage Probability}

The OP can be defined as the probability that the instantaneous mutual information of the ERA scheme drops below a target spectral efficiency (SE) threshold $R_{\rm th}$ [b/s/Hz], i.e., $P_{\rm out}^\era = \Pr (\log_2(\mathrm{SNR}^\era + 1) < R_{\rm th})$.
From \eqref{SNR_ERA}, and invoking Theorem \ref{theorem_dist_Z_Gamma}, an approximate closed-form expression for the OP of the ERA scheme can be obtained as 
\begin{align}
	P_{\rm out}^{\era, \mathrm{Gam}}
	= \Pr \left( \avgSNR Z^2 \leq \SNRth \right) 
	\overset{\rm (a)}{\approx} F_{Z} \left(\sqrt{\SNRth / \avgSNR }\right), 
	\label{eq_ERA_Gamma_OP_end}
\end{align}
where $\SNRth = 2^{R_{\rm th}} - 1$ and step (a) in \eqref{eq_ERA_Gamma_OP_end} is due to the fact that $F_{X^2}(x) = F_{X} (\sqrt{x}), x>0$.

\subsubsection{Ergodic Capacity}

The system ergodic capacity can be determined by averaging the system instantaneous capacity over a large number of channel realizations. 
Invoking Theorem \ref{theorem_dist_Z_Gamma}, an approximate closed-form expression for the EC of the ERA scheme can be expressed as
\begin{align}
	&\bar{C}^{\era, \mathrm{Gam}}
	= \mathbb{E} \left[ \log_2 \left( 1 + \mathrm{SNR}^\era \right) \right] \nonumber \\
	&\quad= \mathbb{E} \left[ \log_2 \left( 1 + \avgSNR Z^2 \right) \right] \nonumber \\
	&\quad\approx \frac{1}{\Gamma({\alpha_\idz}) \ln 2 } \frac{2^{{\alpha_\idz} -1}}{\sqrt{\pi}}
		G^{5,1}_{3,5} \left( \frac{ ({\beta_\idz})^2 }{4  \avgSNR} \middle\vert
		\begin{matrix}
		0, \frac{1}{2}, 1 \\ \frac{{\alpha_\idz}}{2}, \frac{{\alpha_\idz} +1}{2},0,\frac{1}{2},0
		\end{matrix} \right) .   
	\label{eq_ERA_Gamma_EC_end}
\end{align}
The detailed derivation of \eqref{eq_ERA_Gamma_EC_end} is provided in Appendix \ref{appx_ec_ctt_Gamma}.

\subsection{Statistical Channel Characterization of the ERA Scheme Based on Log-Normal Distribution}\label{subsec_LogNormal_Z}

To give more insights into the ERA scheme's fading model, we show that the Log-Normal distribution can alternatively be used to approximate the true distribution of the magnitude of the e2e channel coefficient of the ERA scheme.

\begin{Theorem} \label{theorem_dist_Z_LogNormal}
	The true distribution of $Z$ can be approximated by the Log-Normal distribution, which is characterized by two parameters $\nu_\idz$ and $\zeta_\idz$, i.e., $Z \overset{\rm approx.}{\sim} \mathrm{LogNormal} (\nu_\idz, \zeta_\idz)$, where the estimators of $\nu_\idz$ and $\zeta_\idz$ are given in \eqref{para_log_normal_nu_Z} and \eqref{para_log_normal_zeta_Z}, respectively.
\end{Theorem}

\begin{IEEEproof}
Since the Log-Normal distribution is characterized by two parameters $\nu$ and $\zeta$, following Step 2 of our framework, we need to match the first two moments of $Z$, i.e., solving the following system of equations: $\mathbb{E}[Z] = \mathbb{E}[X]$ and $\mathbb{E} [Z^2] = \mathbb{E}[X^2]$, where $X$ is a Log-Normal RV. As a result, the estimators of $\nu_\idz$ and $\zeta_\idz$ can be written as \cite{Fenton_TCOM_1960}
\begin{align}
	\nu_\idz &= 
	\ln \left( \frac{ \left( \mathbb{E}[Z] \right)^2 }
	{ \sqrt{ \mathbb{E}[(Z)^2] } } 
	\right)
	=
	\ln \left( \frac{ \left( \mu_\idz (1) \right)^2 }
	{ \sqrt{ \mu_\idz (2) } } 
	\right)
	, \label{para_log_normal_nu_Z} \\
	\zeta_\idz &=
	\sqrt{ \ln \left( 
		\frac{  \mathbb{E}[Z^2] }{  (\mathbb{E}[Z])^2 } 
		\right) } 
	= \sqrt{ \ln \left( 
		\frac{  \mu_\idz (2) }{  (\mu_\idz (1))^2 } 
		\right) }
	. \label{para_log_normal_zeta_Z}
\end{align}
Next, in order to determine the exact values of $\nu_\idz$ and $\zeta_\idz$, we derive a general expression for the $k$-th moment of $Z$, i.e., $\mu_\idz (k) = \mathbb{E}[Z^k]$. After some mathematical manipulations, $\mu_\idz (k)$ can be obtained as
\begin{align}
	\mu_\idz (k)  &= \sum_{l=0}^{k} \binom{k}{l} \mathbb{E}[(h_0)^l] \mathbb{E}[T^{k-l}] \nonumber \\
	&= \sum_{l=0}^{k} \binom{k}{l} \mu_{h_0} (l) \mu_T (k-l),
\end{align}
where $\mu_{h_0} (k)$ and $\mu_T (k)$ are derived in \eqref{k_moment_h0} and \eqref{mu_T_k}, respectively.
Consequently, the PDF, $f_{Z} (x; \nu_\idz, \zeta_\idz)$, and the CDF, $F_{Z} (x;, \nu_\idz, \zeta_\idz)$, can be expressed using \eqref{PDF_log_normal} and \eqref{CDF_log_normal}, respectively. This completes the proof of Theorem \ref{theorem_dist_Z_LogNormal}.
\end{IEEEproof}

\subsubsection{Outage Probability}

An approximate closed-form expression for the OP of the ERA scheme can be obtained as
\begin{align}
	P_{\rm out}^{\era, \rm LN}
	&= \Pr \left( \avgSNR Z^2 \leq \SNRth \right) \nonumber \\
	&= F_{Z} \left( \sqrt{\frac{\SNRth}{\avgSNR}} \right) \nonumber \\
	&\approx 
	\frac{1}{2} + \frac{1}{2} \mathrm{erf} \left( \frac{ \ln \left( \sqrt{\frac{\SNRth}{\avgSNR}} \right) - \nu_\idz }{ \sqrt{2 \zeta_\idz^2} } \right). 
	\label{eq_ERA_LogNormal_OP_end}
\end{align}

\subsubsection{Ergodic Capacity} \label{subsubsec_ERA_LN_EC}
The EC of the ERA scheme can be determined as $\bar{C}^{\era,\mathrm{LN}} = \mathbb{E} \left[ \log_2 \left( 1 + \avgSNR Z^2 \right) \right]$.

\begin{Lemma} \label{lemma_Upsilon}
Given $X \sim \mathrm{LogNormal}(\nu_\idx, \zeta_\idx)$, and let $\bar{C}_X = \mathbb{E} \left[ \log_2 \left( 1 + \avgSNR X \right) \right]$, it can be shown that $\bar{C}_X  = \Upsilon(\nu_\idx, \zeta_\idx)$, where
\begin{align} \label{eq_Upsilon_end}
	\Upsilon (\nu_\idx, \zeta_\idx) &\triangleq \frac{1}{\ln(2)} \bigg[ \Xi \left( \frac{1}{\zeta \sqrt{2}}, \frac{\ln(\avgSNR) + \nu}{\zeta \sqrt{2}}\right) \nonumber \\
	&\quad+ \Xi \left( \frac{1}{\zeta \sqrt{2}}, - \frac{\ln(\avgSNR) + \nu}{\zeta \sqrt{2}}\right) + \frac{\zeta}{\sqrt{2 \pi}} e^{ - \frac{(\ln(\avgSNR) + \nu)^2}{2 \zeta^2}} \nonumber \\
	&\quad+ \frac{\ln(\avgSNR) + \nu}{2} \mathrm{erfc} \left( - \frac{\ln(\avgSNR) + \nu}{\zeta \sqrt{2}} \right) \bigg].
\end{align} 
Here, $\Xi(\cdot,\cdot)$ is presented in \eqref{Xi_a} and derived in \eqref{eq_Xi_end}.
\end{Lemma}

\begin{IEEEproof}
We have
\begin{align}
	\bar{C}_X &= \mathbb{E} \left[ \log_2 \left( 1 + \avgSNR X \right) \right]  
	= \int_{0}^{\infty} \log_2 (1 + x) \frac{1}{\avgSNR} f_X \left( \frac{x}{\avgSNR}\right) dx . \label{ec_ln_x_a}
\end{align}
Substituting \eqref{PDF_log_normal} into \eqref{ec_ln_x_a} yields
\begin{align}
	\bar{C}_X &= \frac{1}{\avgSNR \zeta \sqrt{2 \pi}} \int_{0}^{\infty} \frac{\log_2(1+x)}{x/\avgSNR}
	e^{ - \frac{(\ln(x/\avgSNR) - \nu)^2}{2 \zeta^2} } dx.
\end{align}
Since $\log_2(1+x) = \ln(1+x) / \ln(2)$, we have 
\begin{align} \label{eq_C_X_b}
	\bar{C}_X &= \frac{1}{\zeta \sqrt{2 \pi} \ln(2)} 
	\int_{0}^{\infty} \frac{\ln(1+x)}{x}
	e^{ - \frac{[\ln(x) - (\ln(\avgSNR) + \nu)]^2}{2 \zeta^2} } dx.
\end{align}
Let $\Xi(\cdot,\cdot)$ be defined as
\begin{align} \label{Xi_a}
	\Xi(a,b) 
	\triangleq 
	\frac{a}{\sqrt{\pi}} \int_0^{1} \frac{\ln(1+x)}{x} e^{- [a\ln(x) - b]^2} dx .
\end{align}
From the fact that $\ln(1+x) = \sum_{k=1}^8 { a_k x^k + \epsilon(x) },~ 0 < x < 1$ \cite{Laourine_LCOMM_2007}, next, using the identity \cite[Eq.~(8.252.6)]{Gradshteyn2007}, and then using the fact that $1-{\rm erf}(x) = {\rm erfc}(x)$ \cite[Eq.~(8.250.4)]{Gradshteyn2007}, after some mathematical manipulations, $\Xi(a,b)$ in \eqref{Xi_a} can be approximated by
\begin{align} \label{eq_Xi_end}
    \Xi(a,b) \approx \frac{e^{b^2}}{2}
        \sum_{k=1}^{8} a_k
        \exp\left( \left( b+\frac{k}{2a} \right)^2 \right)
        {\rm erfc}\left( b+\frac{k}{2a} \right).
\end{align}
Using \eqref{eq_Xi_end}, and following \cite[Subsection II-A]{Laourine_LCOMM_2007}, one can arrive at the approximate closed-form expression for $\Upsilon (\nu_\idx, \zeta_\idx)$ in \eqref{eq_Upsilon_end}. This completes the proof of Lemma \ref{lemma_Upsilon}.
\end{IEEEproof}

Based on $Z \overset{\rm approx.}{\sim} \mathrm{LogNormal}(\nu_\idz, \zeta_\idz)$ as stated in Theorem \ref{theorem_dist_Z_LogNormal}, we can derive the PDF of $Z^2$, however, the resultant PDF of $Z^2$ using this method leads to a non-closed-form expression of the integral in \eqref{ec_ln_x_a}. To circumvent this problem, we perform Step $2$ of the proposed statistic characterization framework again, and after some mathematical manipulation steps, we obtain $Z^2 \overset{\rm approx.}{\sim} \mathrm{LogNormal}(\nu_{\idz^2}, \zeta_{\idz^2})$, where $\nu_{\idz^2} = \ln \left( \frac{ \left( \mu_\idz (2) \right)^2 }{ \sqrt{ \mu_\idz (4) } } \right)$, $\zeta_{\idz^2} = \sqrt{ \ln \left( \frac{  \mu_\idz (4) }{  (\mu_\idz (2))^2 } \right) }$.
From \eqref{ec_ln_x_a}, an approximate closed-form expression for the EC of the ERA scheme can be obtained as
\begin{align}
	\bar{C}^{\era,\rm LN}
	= 
	\mathbb{E} \left[ \log_2 \left( 1 + \avgSNR Z^2 \right) \right] 
	\approx 
	\Upsilon(\nu_{\idz^2}, \zeta_{\idz^2}). 
	\label{eq_ERA_LogNormal_EC_end}
\end{align}

\section{Performance Analysis of the ORA Scheme}\label{section_analysis_ORA}

\subsection{Statistical Channel Characterization of the ORA Scheme}\label{subsection_ORA_Gamma}

We first rely on the proposed framework to match the true distribution of the key component of $R$, i.e., $V_n$, to a Gamma distribution, as stated in \eqref{eq_Gamma_V_n}, and then we derive approximate closed-form expressions for the PDF and CDF of $R$.
Specifically, since $h_0$ and $V_n$ are independent RVs and $h_0, V_n \geq 0$, the opportunistic scheduling criterion in \eqref{selection_criterion_a} can be re-expressed as
\begin{align}
	n^* = \arg \max_{1\leq n \leq N} \left\{ \avgSNR \left[ h_0 + V_n \right]^2 \right\}
	= \arg \max_{1\leq n \leq N}  V_n. 
	\label{selection_criterion_b}
\end{align}
Thus, the e2e SNR of the ORA scheme in  \eqref{SNR_ORA_n} can be rewritten as 
\begin{align}
	\mathrm{SNR}^\ORA_{n^*} 
	= 
	\avgSNR \left[ h_0 + M_V \right]^2 .
	\label{SNR_ORA_n_star}
\end{align}
The statistical characterization of the magnitude of the e2e channel coefficient of the ORA scheme is presented in the following theorem.

\begin{Theorem} \label{theorem_dist_R_Gamma}

	Approximate closed-form expressions for the CDF and PDF of $R$ can be, respectively, obtained as
	\begin{align} 
		F_{R} (x) 
		\approx
		\Phi(x) 
		&\triangleq
		\sum_{m=1}^M 
		\left[ F_{h_0} \left( \frac{m}{M} x \right) 
		- 
		F_{h_0} \left( \frac{m-1}{M} x \right) \right] \nonumber \\
		&\quad\times F_{M_V} \left( \frac{M-m+1}{M} x \right).  \label{CDF_R_end}
	\end{align}
	\begin{align}
		&f_R (x) 
		\! \approx \!
		\Delta (x) 
		\! \triangleq \!
		\sum_{m=1}^M 
		\bigg[
		\bigg[
		\frac{m}{M} f_{h_0} \left( \frac{m}{M} x \right)
		\!-\! \frac{m-1}{M} f_{h_0} \left( \frac{m-1}{M} x \right) \!
		\bigg] \nonumber \\
		& \times
		F_{M_V} \left( \frac{M-m+1}{M} x \right) \frac{M-m+1}{M}
		\nonumber \\
		&\times 
		\left( 
		F_{h_0} \left( \frac{m}{M} x \right)
		- F_{h_0} \left( \frac{m-1}{M} x \right) 
		\right) f_{M_V} \left( \frac{M-m+1}{M} x \right)
		\bigg] . \label{PDF_R_end}
	\end{align}
	where $f_{h_0} (\cdot)$, $F_{h_0} (\cdot)$, and $F_{M_V} (\cdot)$ are presented in \eqref{PDF_Naka}, \eqref{CDF_Naka}, and \eqref{CDF_MV_end}, respectively.
\end{Theorem}

\begin{IEEEproof}
Since the $\max \{ \cdot \}$ operation on $V_n$ changes the statistical characteristics of its argument, $V_n$, we first rely on the Total Probability Theorem \cite{Peebles2000} and the definition of conditional probability \cite{Peebles2000} to determine the probability distribution of $R$. 
The CDF of $R$, $F_R (x)$, can be expressed as 
\begin{align}
	F_R (x) 
	= \Pr (R \leq x) 
	= \Pr ( (h_0 + M_V) \leq x) .
	\label{CDF_R_a}
\end{align}
From \eqref{selection_criterion_b} and \eqref{SNR_ORA_n_star}, \eqref{CDF_R_a} can be further expressed as
\begin{align} 
  F_R (x) 
  &= \Pr ( (h_0 + V_{n^*}) \leq x) \nonumber \\
  &= \sum_{n=1}^{N}  \Pr ( n^* = n) \Pr \left( (h_0 + V_{n^*}) \leq x \mid n^* = n \right) .
  \label{CDF_R_b}
\end{align}
Based on the conditional probability definition \cite{Peebles2000}, we have
\begin{align} \label{CDF_R_c}
	&\Pr \left( (h_0 + V_{n^*}) \leq x \mid n^* = n \right) \nonumber \\
	&\quad =
	\frac{\Pr (\{(h_0 + V_{n^*}) \leq x \} \cap \{n^* = n\} )}
	{\Pr ( n^* = n) } .
\end{align}
Thus, \eqref{CDF_R_b} can be rewritten as
\begin{align} 
  F_R (x) = \sum_{n=1}^{N}  \Pr (\{(h_0 + V_{n^*}) \leq x \} \cap \{n^* = n\} ) .
  \label{CDF_R_d}
\end{align}
From \eqref{SNR_ORA_n} and \eqref{selection_criterion_b}, \eqref{CDF_R_d} can be further expressed as
\begin{align} 
	F_R (x) 
	= \sum_{n=1}^{N} \Pr ( \{(h_0 + V_{n}) \leq x \} \cap \{V_n = \max_{1\leq k \leq N} V_k \}) .
	\label{CDF_R_e}
\end{align} 
It is noted that $V_n = \max_{1\leq k \leq N} V_k, n=1,\ldots,N,$ are pairwise disjoint events, and 
\begin{align}
	\sum_{n=1}^N \Pr \left( V_n = \max_{1\leq k \leq N} V_k \right) = 1 .
\end{align}
Thus, for an arbitrary event $A$, relying on the Total Probability Theorem \cite{Peebles2000}, we have
\begin{align}
	\sum_{n=1}^N \Pr \left( \left\{ V_n = \max_{1\leq k \leq N} V_k \right\} \cap A \right) = \Pr (A) .
\end{align}
Thus, \eqref{CDF_R_e} can be further expressed as
\begin{align}
	F_R (x)
	&= \Pr \left( \big\{ \max_{1\leq k \leq N} V_k \leq (x - h_0) \big\} \cap \{ h_0 < x \}\right) \nonumber \\
	&= \int_0^x F_{M_V} (x - y) f_{h_0} (y) dy .
	\label{CDF_R_f}
\end{align}
Since $V_k$, $k=1,\ldots,N$, are i.n.i.d. RVs, from \eqref{CDF_Vn_end}, the CDF of $M_V$ can be derived as
\begin{align}
	F_{M_V} (x) 
	&= \Pr \left( \max_{1\leq k \leq N} V_k \leq x \right) \nonumber \\
	&= \prod_{k=1}^N F_{V_k} (x)
	\approx \prod_{k=1}^N 
	\frac{ \gamma(L_k {\alpha_\idu}_k, {\beta_\idu}_k z) }
	{ \Gamma \left( L_k {\alpha_\idu}_k \right) } . 
	\label{CDF_MV_end}
\end{align}
Inserting \eqref{CDF_MV_end} into \eqref{CDF_R_f} results in an integral that cannot be expressed in a closed-form. \label{page_technical_challenge_2} To address this problem, we further express \eqref{CDF_R_f} as 
\begin{align} \label{CDF_R_g}
	F_R (x) = \int_0^x \int_0^{x - y} f_{M_V} (z) f_{h_0}(y) dz dy .
\end{align}
Using the $M$-staircase approximation, \eqref{CDF_R_g} can be derived as
\begin{align}
	F_{R} (x) 
	&\overset{M \to \infty}{\approx} 
	\sum_{m=1}^M 
	\int_{\frac{m-1}{M} x}^{\frac{m}{M} x} f_{h_0} (y) 
	\int_{0}^{\frac{M-m+1}{M} x} f_{M_V} (z) dz dy . \label{CDF_R_k}
\end{align}
After some mathematical manipulations, \eqref{CDF_R_k} can be derived as \eqref{CDF_R_end}. Next, by taking the derivative of \eqref{CDF_R_end}, i.e., $ f_{h_0} (x) = \frac{d F_{h_0}(x)}{dx}$, the PDF of $R$ is obtained as in \eqref{PDF_R_end}, while the PDF of $M_V$, $f_{M_V} (x)$, is derived by taking the derivative of \eqref{CDF_MV_end}. This completes the proof of Theorem \ref{theorem_dist_R_Gamma}.
\end{IEEEproof}

\subsubsection{Outage Probability}

Invoking Theorem \ref{theorem_dist_R_Gamma}, an approximate closed-form expression for the OP of the ORA scheme can be obtained as
\begin{align}
	P_{\rm out}^{\ORA, \rm Gam}
	&= \Pr (\avgSNR R^2 \leq \SNRth) = F_{R^2} \left(\SNRth / \avgSNR \right) \nonumber \\
	&= F_{R} \left(\sqrt{\SNRth / \avgSNR} \right)
	\approx \Phi \left(\sqrt{\SNRth / \avgSNR} \right) , 
	\label{eq_ORA_Gamma_OP_end}
\end{align}
where $\Phi (\cdot)$ is presented in \eqref{CDF_R_end}.

\subsubsection{Ergodic Capacity} 

An analytical closed-form expression for the EC of the ORA scheme can be expressed as
\begin{align}
	\bar{C}^{\ORA, \mathrm{Gam}} 
	&= \mathbb{E}[\log_2 ( 1 + \avgSNR R^2)] \nonumber \\
	&=
	\frac{1}{\ln 2} \int_{0}^\infty 
	\frac{1}{z+1} 
	\left[ 1 - F_{R} \left( \sqrt{\frac{z}{\avgSNR}} \right) \right] dz .
	\label{eq_ORA_Gamma_EC_end}
\end{align}
With the PDF of $R$ in \eqref{CDF_R_end}, the integral in \eqref{eq_ORA_Gamma_EC_end} does not admit a closed-form expression. Nevertheless, an approximate closed-form expression for the EC of the ORA scheme is obtained in subsection \ref{subsection_ORA_ec_log_normal}. \label{page_technical_challenge_1}

\subsection{Statistical Channel Characterization of the ORA Scheme Based on Log-Normal Distribution}

Using the Gamma distribution approximation presented in subsection \ref{subsection_ORA_Gamma}, the closed-form expression for the EC of the ORA scheme cannot be derived. It motivates us to alternatively use a Log-Normal distribution to statistically characterize the e2e channel coefficient.

\begin{Theorem} \label{theorem_dist_R_LogNormal}
	The true distribution of $R$ can be approximated by the Log-Normal distribution, characterized by two parameters $\nu_\idr$ and $\zeta_\idr$, i.e.,
	%
%	\begin{align}
		$R \overset{\rm approx.}{\sim} \mathrm{LogNormal} (\nu_\idr, \zeta_\idr)$,
%	\end{align}
	%
	where the estimators of $\nu_\idr$ and $\zeta_\idr$ are presented in \eqref{para_log_normal_nu_r} and \eqref{para_log_normal_zeta_r}, respectively.
\end{Theorem}

\begin{IEEEproof}
By using the same mathematical reasoning as in Subsection \ref{subsec_LogNormal_Z}, the parameters characterizing the Log-Normal distribution of $R$ can be obtained as
\begin{align}
	\nu_{\idr} &= \ln \left( 
	\frac{ \left( \mathbb{E}[R] \right)^2 }
	{ \sqrt{ \mathbb{E}[R^2] } } 
	\right)
	= \ln \left( 
	\frac{ \left( \mu_\idr (1) \right)^2 }
	{ \sqrt{ \mu_\idr (2) } } 
	\right)
	, \label{para_log_normal_nu_r} \\
	\zeta_{\idr} &= \sqrt{ \ln \left( 
		\frac{  \mathbb{E}[R^2] }{  (\mathbb{E}[R])^2 } 
		\right) } 
	\!=\! \sqrt{ \ln \left( 
		\frac{  \mu_\idr (2) }{  (\mu_\idr (1))^2 } 
		\right) }
	. \label{para_log_normal_zeta_r}
\end{align}
To obtain the value of $\nu_\idr$ and $\zeta_\idr$, we derive the $k$-th moment of $R$, which can be obtained as
\begin{align}
	\mu_{R} (k) = \sum_{v=0}^{k} \binom{k}{v} \mu_{h_0} (v) \mu_{M_V} (k-v) , \label{mu_k_R_end}
\end{align}
From the CDF of $M_V$ in \eqref{CDF_MV_end}, an expression for $\mu_{M_V} (k)$ is derived in the following Lemma.

\begin{Lemma}\label{lemma_k_th_moment_M_V}
A tight approximate closed-form expression for the $k$-th moment of $M_V$ can be obtained as in \eqref{mu_M_V_k_end} at the top of the next page.
\begin{figure*}
\begin{align}
	\mu_{M_V} (k) 
	&=
	\left[ \sum_{t=1}^N {\beta_\idu}_t \right]^{-k}
	\left[ \prod_{t=1}^N \chi_t^{L_t \alpha_{\idu_t} } \right]
	\sum_{n=1}^{N} \frac{ \Gamma(\Lambda + k) }{ \Gamma(L_n \alpha_{\idu_n})} 	 
	\prod_{\substack{t=1 \\ t \neq n}}^N \frac{1}{ \Gamma(L_t {\alpha_\idu}_t + 1) } 
	\nonumber \\
	&\quad\times
	F_{A}^{(N-1)} 
	\bigg[ 
	\Lambda + k ; 
	\underbrace{1,\ldots,1}_{ (N-1) \text{ terms}} ; 
	\underbrace{L_1 {\alpha_\idu}_1 + 1,\ldots,L_N {\alpha_\idu}_N+1}_{\text{exclude } (L_n \alpha_{\idu_n} + 1) } ;
	\underbrace{\chi_1,\ldots,\chi_N}_{\text{exclude } \chi_n} 
	\bigg] .
	\label{mu_M_V_k_end}
\end{align}
\hrule
\end{figure*}

\end{Lemma}

\begin{table*}[t]
	\centering
	\caption{Simulation Parameters}
	
	\begin{tabularx}{.9\linewidth}{|l |X || l |l| }
		\Xhline{1.5\arrayrulewidth}
		\textbf{Parameters} & \textbf{Values} & \textbf{Parameters} & \textbf{Values}\\
		\Xhline{1.5\arrayrulewidth}
		$\src \to \des$ distance, $d_{\src \des}$ [m] & 100& Avg. height of RISs' location, $H$ [m] & 10 \\
		\hline
		Number of RISs, $N$ & $5$
		& Total reflecting elements, $\sum_n L_n$ & $[125,200]$ \\
		\hline
		Amplitude reflection coef., $\kappa_{nl}$ &  $1$ \cite{Bjornson_WCL_2020}
		&Bandwidth, $\rm BW$ [MHz] &$10$ \cite{Bjornson_WCL_2020} \\
		\hline
		Carrier frequency, $f_c$ [GHz]   & $3$ \cite{Bjornson_WCL_2020}
		&
		Nakagami shape parameter, $m$& $\sim \mathcal{U}[2,3]$ \\
		\hline
		CDP at RIS, $\tilde{P}_{nl}$ [mW] & $ 7.8$ \cite{Huang_TWC_2019}
		&
		Target SE, $R_{\rm th}$ [b/s/Hz] & $1$\\
		\hline
		CDP at $\src$, $\tilde{P}_\src$, and $\des$, $\tilde{P}_\des$ [mW] & 10  \cite{Huang_TWC_2019}
		&
		Thermal noise power density [dBm/Hz] & $-174$ \cite{Bjornson_WCL_2020} \\
		\hline
		Transmit power, $P_\src$ [dBm]& $[0,30]$ & RIS's coordinate [m]  & $x_{\ris_n} \sim \mathcal{U}[5,95]$, $y_{\ris_n} \sim \mathcal{U}[1,9]$ \\
		\hline
		Antenna gain $G_{\src}$, $G_\des$, $G_{\ris_n}$ [dB]& 5 \cite{Bjornson_WCL_2020} & Noise figure, $\rm NF$ [dBm] & $10$ \cite{Bjornson_WCL_2020} \\
		\Xhline{1.5\arrayrulewidth}
	\end{tabularx}
	\label{table_simulation}
\end{table*}

\begin{IEEEproof} 
	The proof is provided in Appendix \ref{proof_lemma_k_th_moment_M_V}.
\end{IEEEproof}
Invoking Lemma \ref{lemma_k_th_moment_M_V}, and then substituting \eqref{k_moment_h0} and \eqref{mu_M_V_k_end} into \eqref{mu_k_R_end}, we obtain the closed-form expression for $\mu_{R} (k)$. 
This completes the proof of Theorem \ref{theorem_dist_R_LogNormal}.
\end{IEEEproof}

\subsubsection{Outage Probability}

An approximate closed-form expression for the OP of the ORA scheme can be obtained as
\begin{align}
	P_{\rm out}^{\ORA, \mathrm{LN}}
	&=  \Pr (\avgSNR R^2 \leq \SNRth)
	= F_{R} \left( \sqrt{\frac{\SNRth}{\avgSNR}} \right) \nonumber \\
	&\approx \frac{1}{2} + \frac{1}{2} \mathrm{erf} \left( \frac{ \ln \left( \sqrt{\frac{\SNRth}{\avgSNR}} \right) - \nu_\idr }{ \sqrt{2 \zeta_\idr^2} } \right) . 
	\label{eq_ORA_LogNormal_OP_end}
\end{align}

\subsubsection{Ergodic Capacity} \label{subsection_ORA_ec_log_normal}
Using the same argument made in subsection \ref{subsubsec_ERA_LN_EC}, and after some mathematical manipulations, we obtain $R^2 \overset{\rm approx.}{\sim} \mathrm{LogNormal}(\nu_{\idr^2}, \zeta_{\idr^2})$, $\nu_{\idr^2} = \ln \left( \frac{ \left( \mu_\idr (2) \right)^2 }{ \sqrt{ \mu_\idr (4) } } \right)$, $\zeta_{\idr^2} = \sqrt{ \ln \left( \frac{  \mu_\idr (4) }{  (\mu_\idr (2))^2 } \right) }$.		
Invoking Lemma \ref{lemma_Upsilon}, an approximate closed-form expression for the EC of the ORA scheme can be expressed as
\begin{align}
	\bar{C}^{\ORA, \mathrm{LN}}
	= \mathbb{E} \left[ \log_2 \left( 1 + \avgSNR R^2 \right) \right]
	\approx \Upsilon(\nu_{\idr^2}, \zeta_{\idr^2}) .
	\label{eq_ORA_LogNormal_EC_end}
\end{align}

\section{Results and Discussions} \label{section_results}

In this section, we provide representative results on the PDFs and CDFs of $Z$ and $R$, as well as the OP, EC, and energy efficiency (EE) of the ERA and the ORA schemes, not only to verify the correctness of the developed analysis, but also to give insights into the performance of the two schemes. It is noted that the results in this section are reproducible using our Matlab code\footnote{The source code is available at \url{https://github.com/trinhudo/Multi-RIS}}, in which we detail the coding of the analytical results, the obtained CDFs and PDFs, and especially the Lauricell function Type-A, $F_A(\cdot)$, used in \eqref{mu_M_V_k_end}. One of the technical contributions of our code is that we use symbolic computations in Matlab \cite{Matlab2021a} to obtain highly accurate results, e.g., using variable-precision arithmetic ($\mathrm{vpa}$ functions) to obtain analytical results with $32$ significant decimal digits of precision. The reason of using symbolic computations is that the analysis of multi-RIS-aided systems under i.n.i.d. fading channels involves a high number of RVs, especially, when the number of elements of each RIS is large, some non-symbolic operations of floating-point arithmetic in Matlab are invalid and return not-a-number (NaN) results. 

Unless otherwise specified, the simulation parameters are set to the values provided in Table \ref{table_simulation}, where the equivalent noise power at $\des$ is $\sigma_\idd^2 = N_0 + 10 \log(\mathrm{BW}) + \mathrm{NF}$ [dBm]. Considering non-line-of-sight (NLoS) condition in the 3GPP Urban Micro (UMi) path-loss model \cite{3GPP,Yildirim_TCOM_2020}, and \cite{Bjornson_WCL_2020}, the path-loss is integrated in the spread parameter of the Nakagami-$m$ distribution as $\Omega_{\rm XY} = G_\mathrm{X} + G_\mathrm{Y} - 22.7 - 26 \log(f_c) - 36.7 \log(d_{\rm XY}/d_0)$ [dB], where $d_{\rm XY}$ [m] is the distance in two-dimensional Cartesian coordinate system, where $\rm X \in \{ \src, \ris_1, \ldots, \ris_N \}$, $\rm Y \in \{ \ris_1, \ldots, \ris_N, \des \}$, and $d_0$ [m] is the reference distance, herein, $d_0 = 1$ m. Let $\vec{L} \in \mathbb{R}^{1 \times N}$ denote a vector that indicates the number of reflecting elements of each one of the $N$ RISs, i.e., $\vec{L} = [L_n:n=1,\ldots,N]$. Specifically, we consider four reflecting element setting as follows: $\vec{L}_1 = [25, 25, 25, 25, 25]$, is used in Figs.~2-4, $\vec{L}_2 = [40, 40, 40, 40, 40]$, $\vec{L}_3 = [20, 30, 40, 50, 60]$, and $\vec{L}_4 = [60, 50, 40, 30, 20]$, which are used in Figs. 5-7. Let $\vec{D} \in \mathbb{R}^{1 \times 2N} = [(x_{\ris_n},y_{\ris_n}): n=1,\ldots,N]$ represent the coordinates of the RISs. The locations of the source and the destination are set as $(0,0)$ and $(100,0)$, respectively, while we consider two location settings for RISs, namely $\vec{D}_1 = [(7,2), (13,6), (41,8), (75,4), (93,3)]$, used in Figs.~2-7, and $\vec{D}_2 = [(5,2), (13,7), (37,6), (69,1), (91,3)]$, used in Fig.~7. \label{page_number_M}The number of steps, $M$, used in the $M$-staircase approximation, is set as $M=100$.

\begin{figure}[h]
	\centering
	\subfloat[]{%
		\includegraphics[width=.9\linewidth]{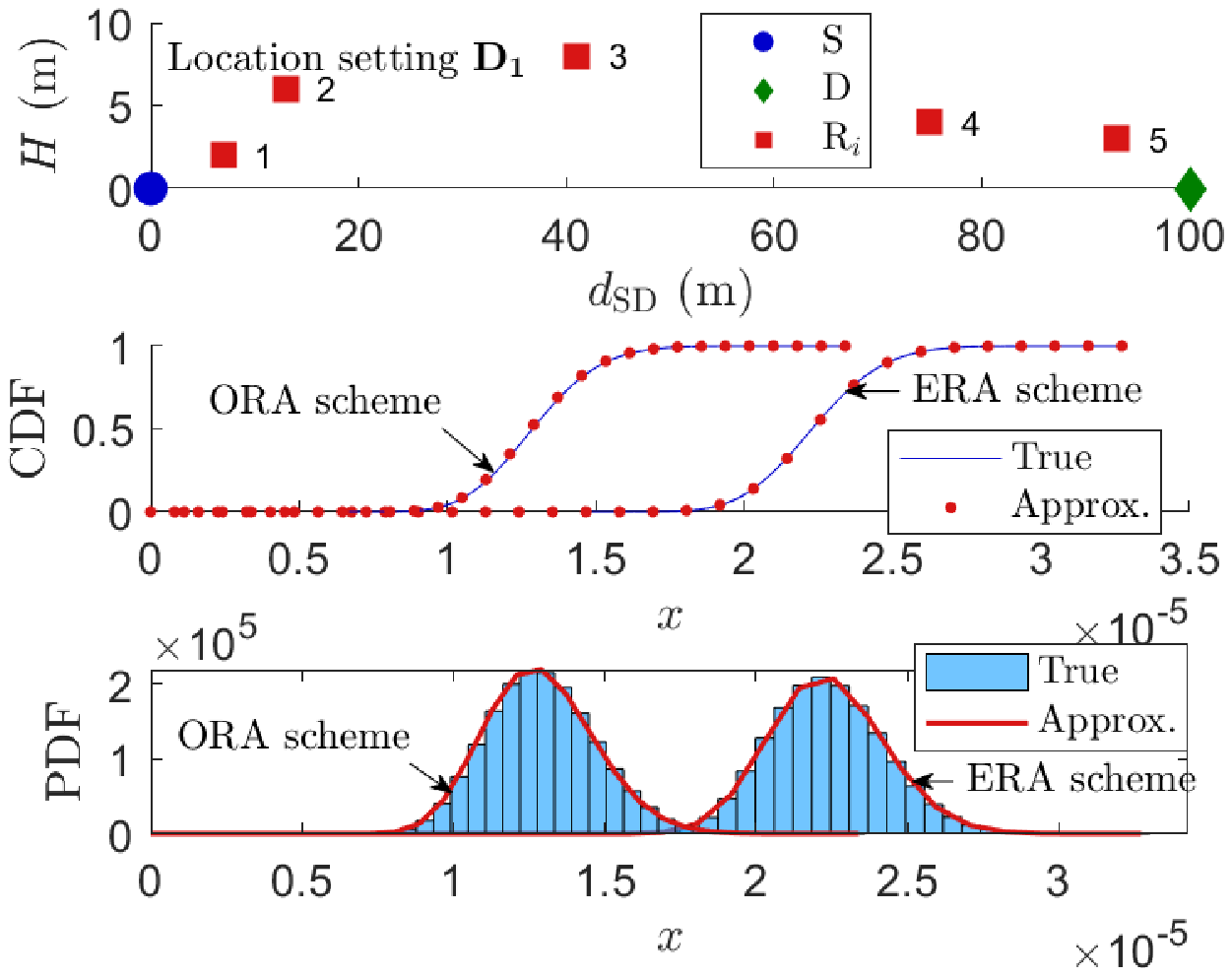}
		\label{fig_CDF_PDF_Gamma}} \hfill
	\subfloat[]{%
		\includegraphics[width=.9\linewidth]{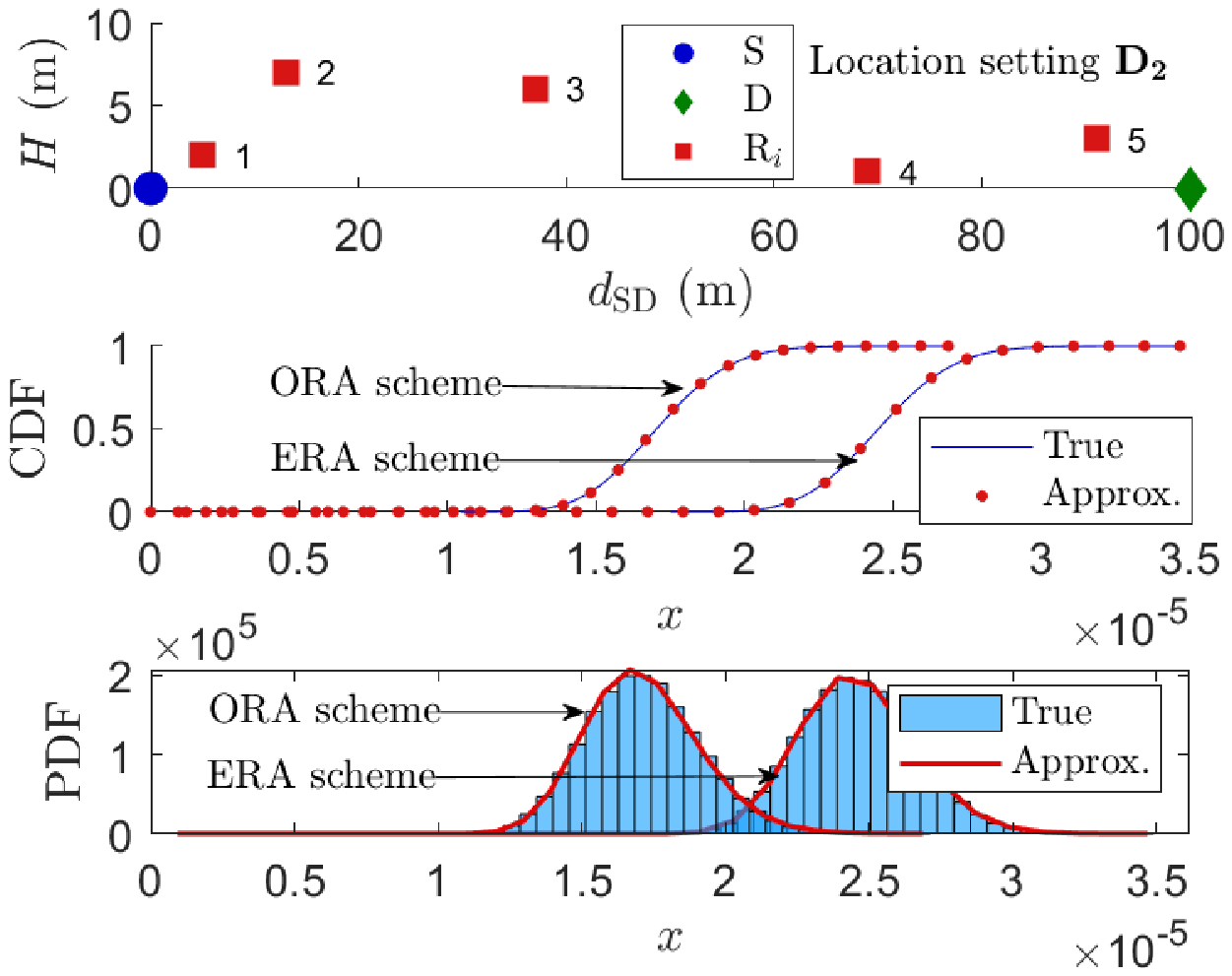}
		\label{fig_CDF_PDF_LN}}
	\caption{Graphical demonstration of the estimated true distributions and the obtained approximate distributions in (a) Theorems \ref{theorem_dist_Z_Gamma} and \ref{theorem_dist_R_Gamma} and (b) Theorems \ref{theorem_dist_Z_LogNormal} and \ref{theorem_dist_R_LogNormal}.}
	\label{fig_CDF_PDF} 
\end{figure}

\begin{figure}[t]
	\centering
	\subfloat[]{%
		\includegraphics[width=.4\linewidth]{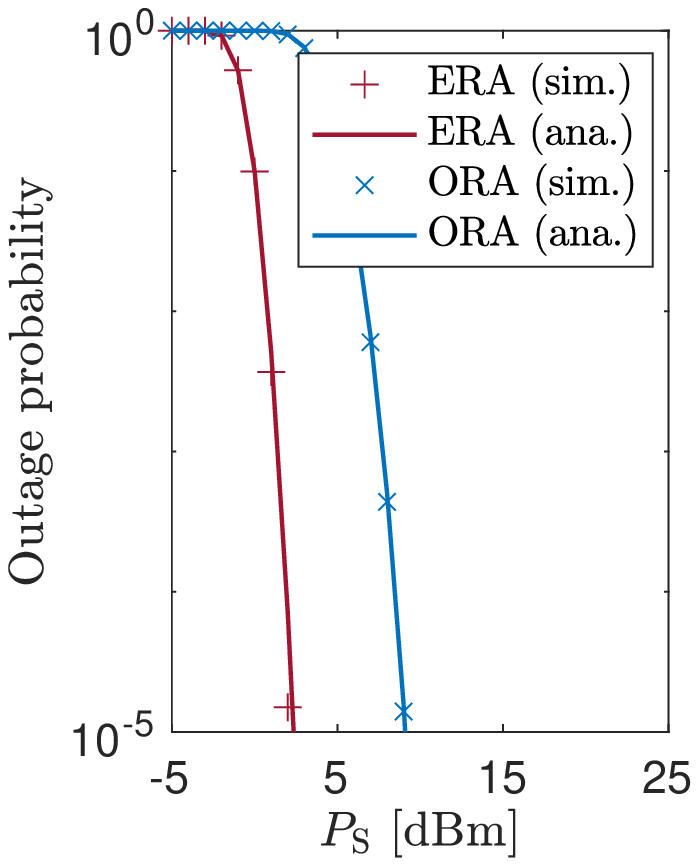}
		\label{fig_OP_EC_a}} 
	\subfloat[]{%
		\includegraphics[width=.39\linewidth]{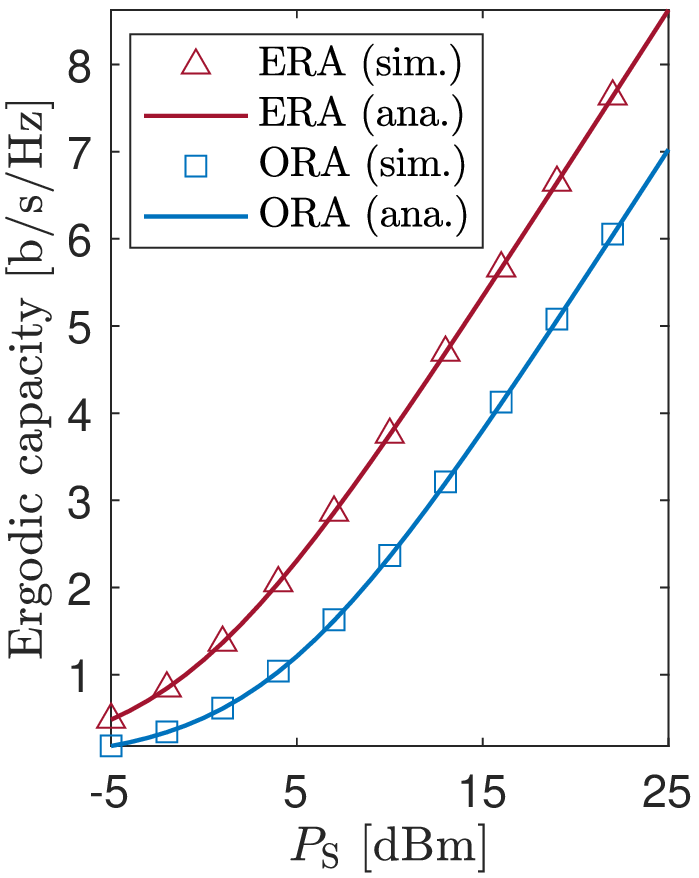}
		\label{fig_OP_EC_b}}  \\
	\subfloat[]{%
		\includegraphics[width=.4\linewidth]{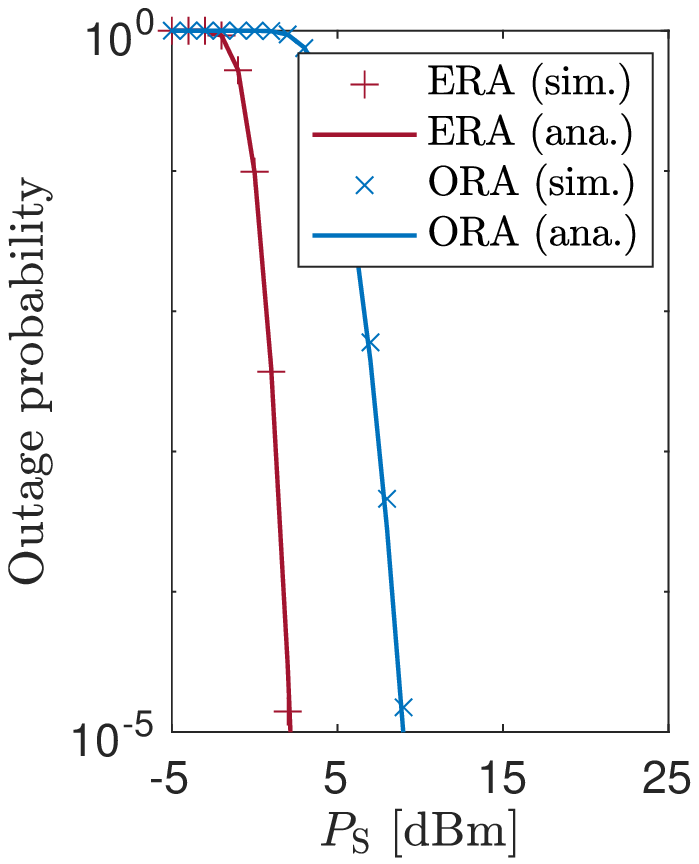}
		\label{fig_OP_EC_c}} 
	\subfloat[]{%
		\includegraphics[width=.39\linewidth]{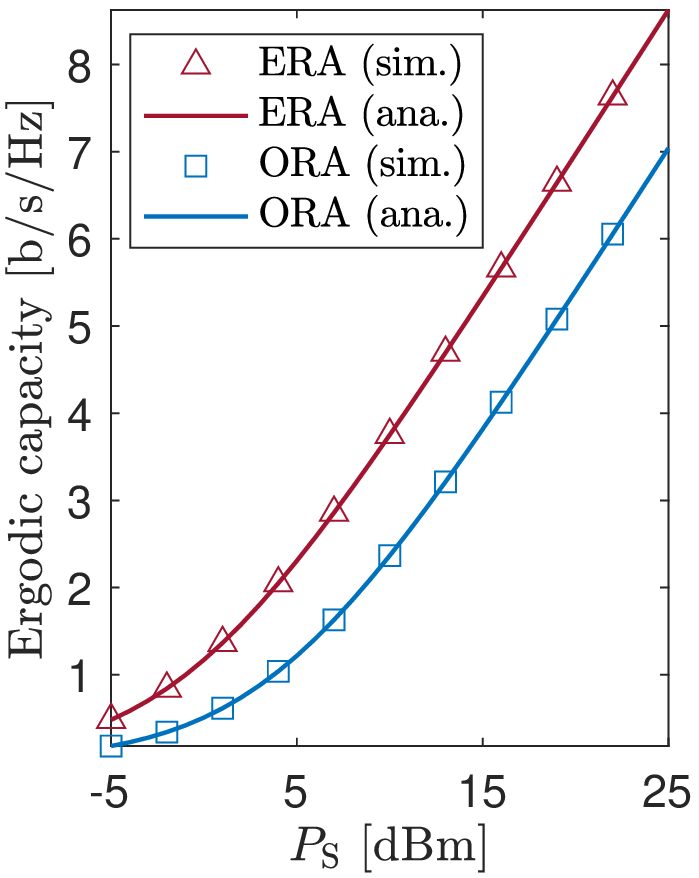}
		\label{fig_OP_EC_d}} 
	\caption{The OP and EC of the ERA and the ORA schemes as a function of transmit power of $\src$, $P_\src$ [dBm], in particular, (a) $P_{\rm out}^{\era, \mathrm{Gam}}$ and $P_{\rm out}^{\ORA, \rm Gam}$, (b) $\bar{C}^{\era, \mathrm{Gam}}$ and $\bar{C}^{\ORA, \mathrm{Gam}}$, (c) $P_{\rm out}^{\era, \rm LN}$ and $P_{\rm out}^{\ORA, \mathrm{LN}}$, and (d) $\bar{C}^{\era, \rm LN}$ and $\bar{C}^{\ORA, \mathrm{LN}}$.}
	\label{fig_verification} 
\end{figure}

\begin{figure*}
	\centering
	\subfloat[]{%
		\includegraphics[width=.25\linewidth]{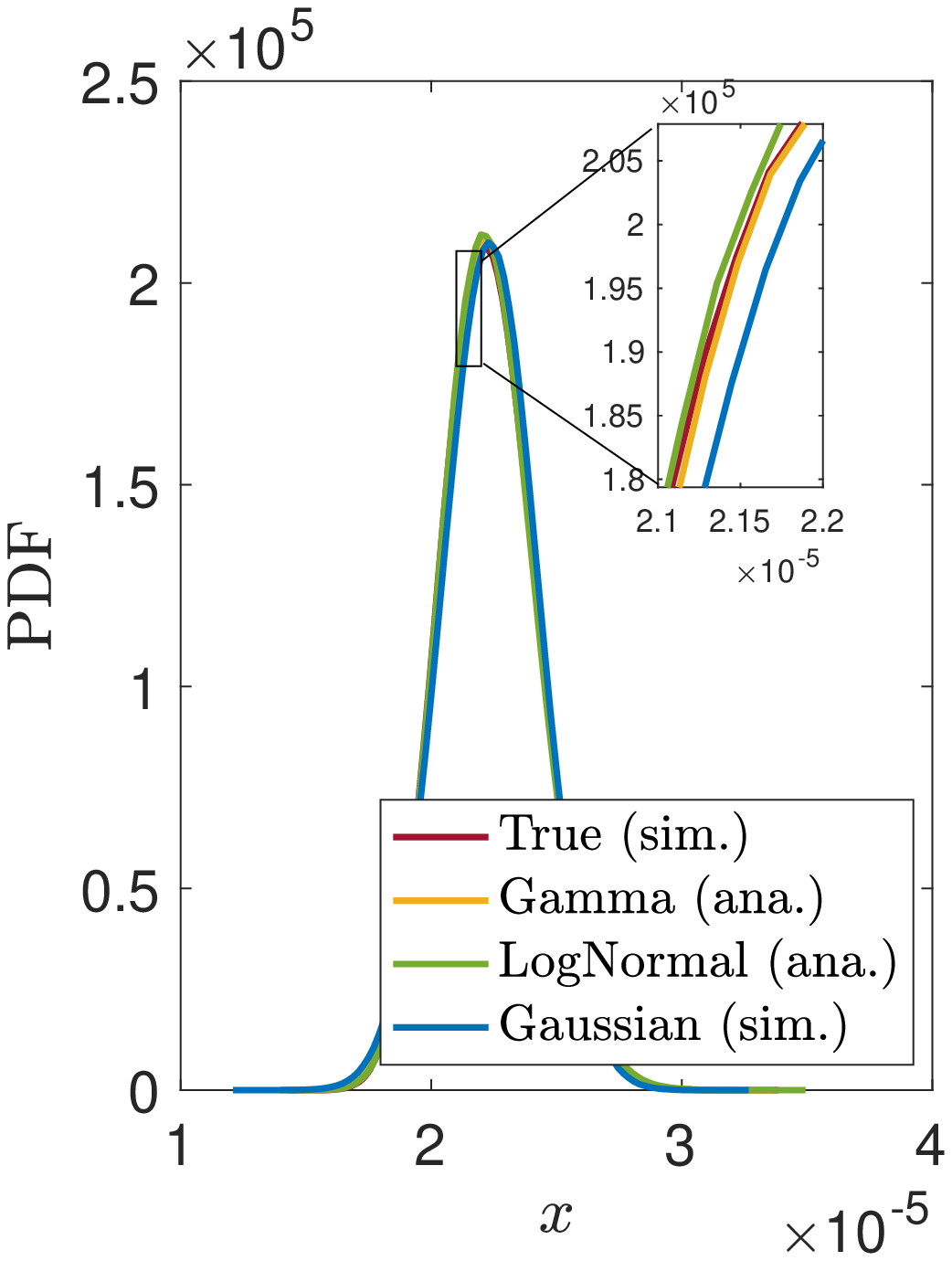}
		\label{fig_4a}} 
	\subfloat[]{%
		\includegraphics[width=.24\linewidth]{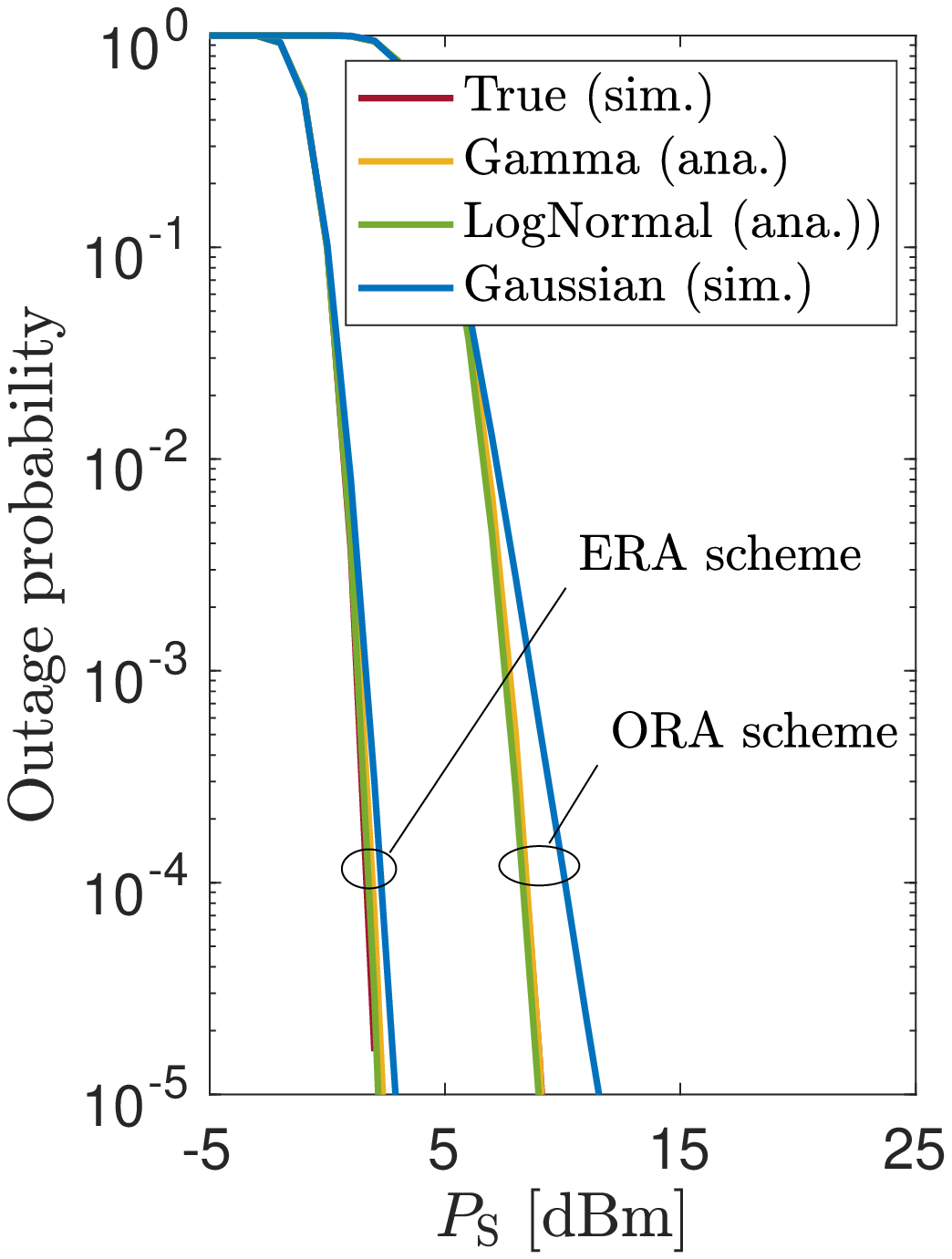}
		\label{fig_4b}} 
	\subfloat[]{%
		\includegraphics[width=.23\linewidth]{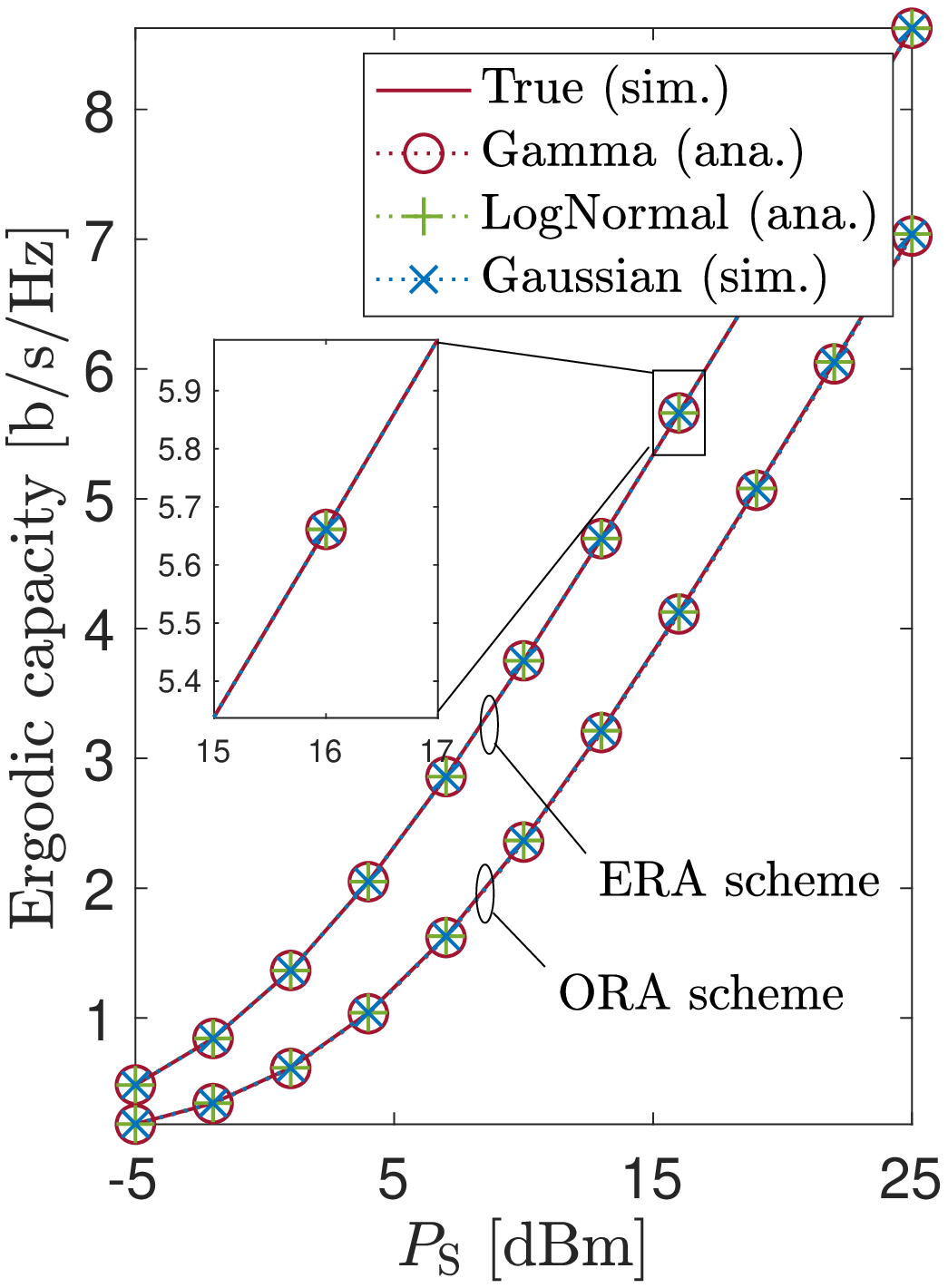}
		\label{fig_4c}} 
	\caption{Comparison between the Gamma, Log-Normal, and Gaussian distribution approximations in terms of a) PDF in the ERA scheme, b) the discrepancy in OP, and c) the discrepancy in EC.}
	\label{fig_distributions} 
\end{figure*}

In Figs.~\ref{fig_CDF_PDF}-\ref{fig_distributions}, we verify the correctness of our developed analysis.
Specifically, in Fig.~\ref{fig_CDF_PDF}, we plot the approximate Gamma and Log-Normal distributions in the ERA and ORA schemes, which are presented in Theorems \ref{theorem_dist_Z_Gamma}-\ref{theorem_dist_R_LogNormal}. As can be seen in Fig.~\ref{fig_CDF_PDF}, the approximate CDFs and PDFs are in agreement with the true CDF and PDF, which are estimated numerically based on simulation data. Next, in Fig.~\ref{fig_verification}, we plot the OP and EC results of the ERA and the ORA schemes, i.e., 
$P_{\rm out}^{\era, \mathrm{Gam}}$,
$\bar{C}^{\era, \mathrm{Gam}}$,
$P_{\rm out}^{\era, \rm LN}$,
$\bar{C}^{\era, \rm LN}$,
$P_{\rm out}^{\ORA, \rm Gam}$,
$\bar{C}^{\ORA, \mathrm{Gam}}$,
$P_{\rm out}^{\ORA, \mathrm{LN}}$,
and $\bar{C}^{\ORA, \mathrm{LN}}$ presented in
\eqref{eq_ERA_Gamma_OP_end},
\eqref{eq_ERA_Gamma_EC_end},
\eqref{eq_ERA_LogNormal_OP_end},
\eqref{eq_ERA_LogNormal_EC_end},
\eqref{eq_ORA_Gamma_OP_end},
\eqref{eq_ORA_Gamma_EC_end},
\eqref{eq_ORA_LogNormal_OP_end},
and \eqref{eq_ORA_LogNormal_EC_end}, respectively. As can be observed in Fig.~\ref{fig_verification}, the theoretical and simulation results are well corroborated, which validates our developed analysis.

Furthermore, in Fig.~\ref{fig_distributions}, we discuss the performance behavior, and compare the impact of the Gamma, Log-Normal, and Gaussian distributions on the OP and EC.
First, as can be seen in Fig.~(4a), the Gamma and Log-Normal distributions provide better approximations compared to the Gaussian distribution. Moreover, from Fig.~(4b), one can see that, compared to the Gamma and Log-Normal distributions, the OP associated with the Gaussian distribution shows a tangible gap between the approximate and true values at the low OP range, i.e., when the OP is lower than $10^{-2}$. The reason behind this OP discrepancy is the inaccurate approximation occurring at the left-hand tail of the distributions.
Fig.~(4c) shows that the analytical and simulation results of the EC obtained by the Gamma, Log-Normal, or Gaussian distributions are well corroborated even in the high transmission power regime. This EC behavior is in line with the results reported in \cite{AlAhmadiTWC2010}. In particular, the EC is not as sensitive to the numerical inaccuracy as the OP.
In short, the Gamma and Log-Normal distribution approximations cause a small OP discrepancy in the large SNR region; nevertheless, this discrepancy is negligible compared to the OP discrepancy caused by the Gaussian distribution approximation. Moreover, the discrepancy in the EC caused by the Gaussian, Gamma, or Log-Normal distribution approximations is very small in the large SNR region and can be negligible.

\subsection{The Difference in Lower Tail Between Gamma and Log-Normal Distributions} \label{subsection_Gamma_LN}
In this subsection, we elaborate more on difference between the Gamma and Log-Normal distributions. As shown in Fig.~(4b), when we compare the OP computed based on the Gamma and Log-Normal distributions, there is a very small discrepancy, which can be negligible. Rigorously, this slight difference in OP is due to the difference at the \textit{left tail (the lower tail)} between the two distributions.
Specifically, let $\mathcal{N}(0,1)$ denote a standard Gaussian (Normal) RV. Considering a Gamma RV $Y$ with the PDF in \eqref{PDF_Gamma}, mean $\mathbb{E}[Y] = \alpha / \beta$ and variance $\mathrm{VAR}[Y] = \alpha/\beta^2$. Note that $\beta$ is the rate parameter. Following the central-limit theorem (CLT), when $\alpha$ is sufficiently large, $Y$ can be remodeled as
\begin{align}
Y \sim \frac{\alpha}{\beta} \left[ 1+\frac{\mathcal{N}(0,1)}{\sqrt{\alpha}} \right],~ \alpha \to \infty.
\label{eq:CLTGamma}
\end{align}
Considering a Log-Normal RV $W$ with the PDF in \eqref{PDF_log_normal}, mean $\mathbb{E}[W] = e^{\nu +\zeta^2/2}$ and variance $\mathrm{VAR}[W] = (e^{\zeta^2}-1) e^{2\nu+\zeta^2}$, $W$ can be re-expressed as $W = \exp( \nu + \zeta \mathcal{N}(0,1) )$.
Following the method in \cite{Kosti2005}, to show the relationship between the Gamma and Log-Normal distributions, we match their means and variances as follows: 
\begin{gather}
\left\{ 
\begin{array}{l}
    \exp\left(
        \nu + \frac{\zeta^2}{2}
    \right)  = \frac{\alpha}{\beta}, \\
    \left( \exp\left(
        \zeta^2
    \right) -1 \right) 
    \exp\left(
        2\nu+\zeta^2
    \right) = \frac{\alpha}{\beta^2}.
\end{array}
\right. \label{meanAndVarLogGam}
\end{gather}
Solving \eqref{meanAndVarLogGam} with respect to $\alpha$ and $\beta$ yields
\begin{gather}
    \zeta^2 = \log \left( \frac{\alpha+1}{\alpha} \right), ~
    \nu = \log\left( \frac{\alpha}{\beta}  \frac{\sqrt{\alpha}}{\sqrt{\alpha+1} } \right).
\end{gather}
As $\alpha$ is sufficiently large, we have $\zeta \to \frac{1}{\sqrt{\alpha}}$ and $\nu \to \log\big( \frac{\alpha}{\beta} \big)$. Hence, $W$ can be remodeled as
$W \sim
    \frac{\alpha}{\beta} \left[ 1+\frac{\mathcal{N}(0,1)}{\sqrt{\alpha}} \right],~\alpha \to \infty$.

\begin{figure}[t]
    \centering
    \begin{minipage}{0.45\textwidth}
        \centering
        \includegraphics[width=.9\textwidth]{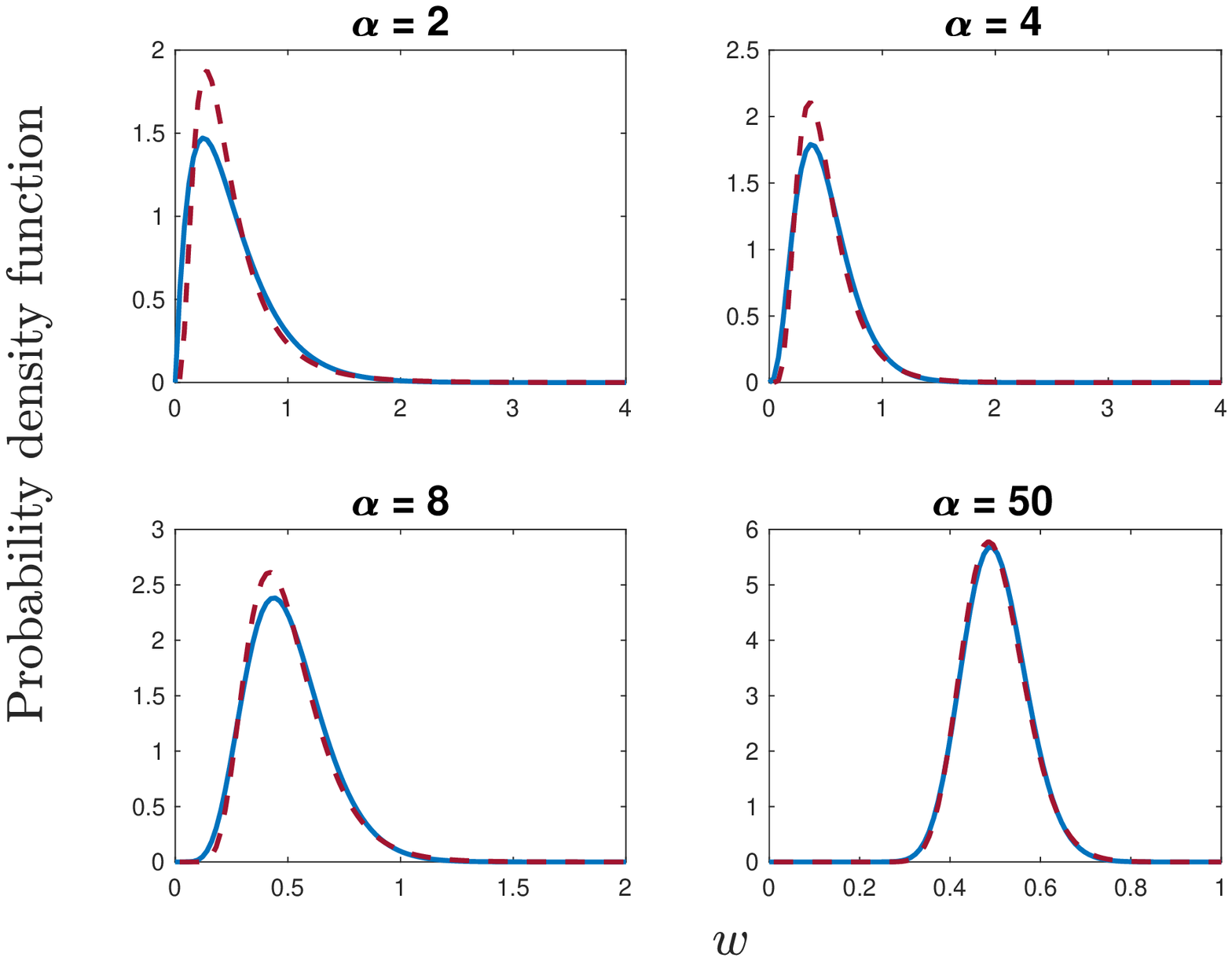}
        \caption{Illustrations of the Gamma PDF (dashed curves) and the Log-Normal PDF (solid curves) with different values of $\alpha$.}
        \label{fig_Gamma_LogNormal}
    \end{minipage}\hfill
    \begin{minipage}{0.45\textwidth}
        \centering
        \includegraphics[width=.8\textwidth]{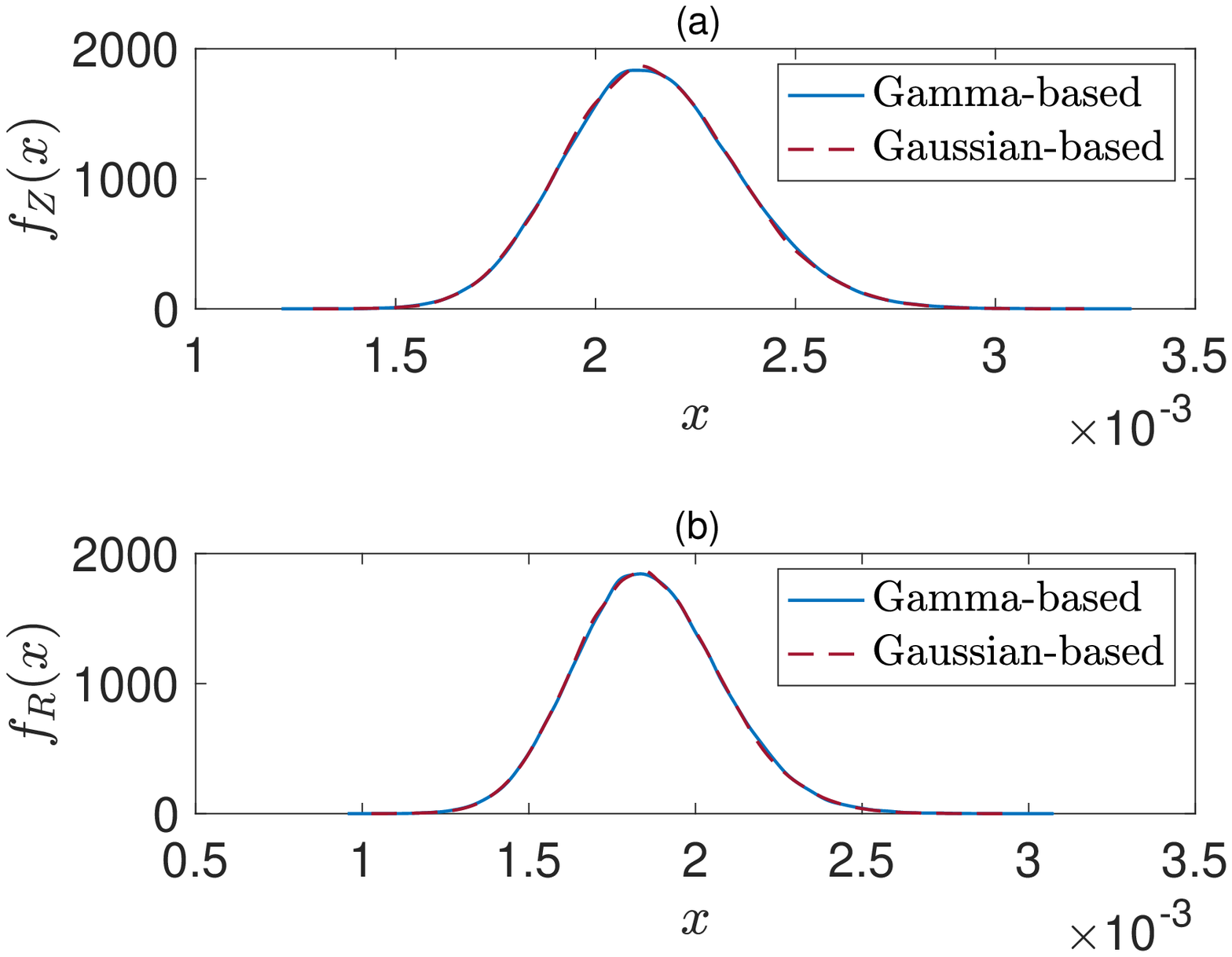} 
        \caption{Comparison between Gamma distribution and the distribution based on the sum of squared Gaussian RVs in (a) the ERA scheme and (b) the ORA scheme.}
        \label{fig_squared_Gaussian}
    \end{minipage}
\end{figure}

This result is consistent with \eqref{eq:CLTGamma}, which shows the similarity between the Gamma and Log-Normal distributions. In Fig. \ref{fig_Gamma_LogNormal}, it can be observed that when $\alpha$ is sufficiently large, e.g., $\alpha \ge 30$, the lower tails of the Gamma and Log-Normal distributions are similar. Note that in our distributed multi-RIS-aided system, the Nakagami shape parameter $m$ of each individual channel is uniformly distributed in the range of $[2,3]$ as in Table~\ref{table_simulation}, consequently, the corresponding shape parameter $\alpha_Z$ in \eqref{alpha_Z} in the ERA scheme is in the range of $[77, 120]$.

\subsection{The Decomposition to The Sum of Squared Normal Random Variables}
\label{page_decomposition}
To give in-depth insight into the physical origin of the e2e channel coefficient of the considered system, we can express the e2e channel coefficients of either the ERA and ORA schemes as the sum of squared Gaussian RVs as follows.

Let $U$, $V$, and $W$ be a Nakagami-$m$ RV, Gamma RV, and Gaussian RV, respectively, specifically, $U \sim \mathrm{Nakagami}(m,\Omega)$, $V\sim \mathrm{Gamma}(\alpha,\zeta)$, and $W \sim \mathcal{N}(0, \sigma^2)$. On one hand, knowing that if $U \sim \mathrm{Nakagami}(m, \Omega)$ and $U = \sqrt{V}$, then $V \sim \mathrm{Gamma}(m, \Omega/m)$. On the other hand, the sum of $M$ squared Gaussian RVs, $\sum_{m=1}^{M} W_m^2$, follows a Gamma distribution, i.e., $\sum_{m=1}^{M} W_m^2 \sim \mathrm{Gamma}(M/2, 2\sigma^2)$.
Based on the above relationship between Nakagami-$m$, Gamma, and Gaussian RVs, assuming that $m_0$, $m_{\idh_n}$ and $m_{\idg_n}$ are integers, we can express the e2e channel coefficient of the ERA and ORA schemes as
\begin{align}
&Z = \sqrt{ \sum_{i=1}^{m_0}\Big( X_{0,i}^2 + Y_{0,i}^2 \Big) } \nonumber \\
&\!\!+\!\! \sum_{n=1}^{N} \sum_{l=1}^{L_n} \! \kappa_{nl}  \!
    \sqrt{ \sum_{j=1}^{m_{{\rm h}_n}}\Big( X_{{\rm h}_n,j}^2 + Y_{{\rm h}_n,j}^2 \Big) } \!
    \sqrt{ \sum_{k=1}^{m_{{\rm g}_n}}\Big( X_{{\rm g}_n,k}^2 + Y_{{\rm g}_n,k}^2 \Big) }, \label{eq_Z_Gauss}\\
&R = \sqrt{ \sum_{i=1}^{m_0} \! \Big( X_{0,i}^2 + Y_{0,i}^2 \Big) }  \nonumber \\
&\!\!+\!\! \max_{1 \le n \le N}
    \sum_{l=1}^{L_n} \kappa_{nl} 
    \sqrt{ \sum_{j=1}^{m_{{\rm h}_n}}\Big( X_{{\rm h}_n,j}^2 \!+\! Y_{{\rm h}_n,j}^2 \Big) } 
    \sqrt{ \sum_{k=1}^{m_{{\rm g}_n}} \!\! \Big( X_{{\rm g}_n,k}^2 \!+\! Y_{{\rm g}_n,k}^2 \Big) }, \label{eq_R_Gauss}
\end{align}
respectively, where $X_{{\rm c},i}$, $Y_{{\rm c},i}$, where ${\rm c} \in \{ 0, {\rm h}_n, {\rm g}_n\}$, are Gaussian RVs with zero means and variances of $\frac{ {\Omega_{\rm c}} }{ 2 m_{\rm c} }$, i.e., $X_{{\rm c},i}, Y_{{\rm c},i} \sim \mathcal{N}\left(0, \frac{ {\Omega_{\rm c}} }{ 2 m_{\rm c} }\right)$. In Figs.~(6a) and (6b), we demonstrate the expressions in  \eqref{eq_Z_Gauss} and \eqref{eq_R_Gauss}, respectively. It is noted that the expressions in \eqref{eq_Z_Gauss} and \eqref{eq_R_Gauss} are valid when the shape parameters of the  Nakagami-$m$ distributions are integer.

\begin{figure}[t]
	\centering
	\subfloat[]{%
		\centering
		\includegraphics[width=.8\linewidth]{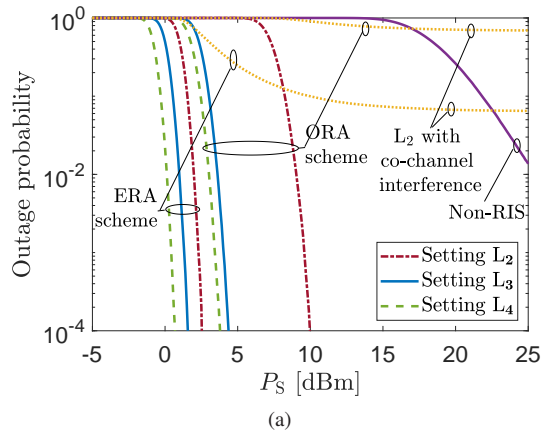}
		\label{fig_OP}} \hspace{2cm}
	\subfloat[]{%
		\includegraphics[width=.8\linewidth]{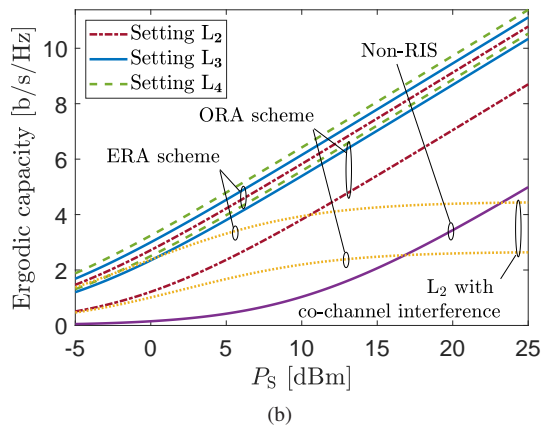}
		\label{fig_EC}}
	\caption{Numerical results demonstrating the impact of the number of reflecting elements on (a) the OP and (b) the EC of the ERA and the ORA schemes, respectively, with co-channel inter-cell interference and $R_{\rm th} = 3$ b/s/Hz.}
	\label{fig_OP_EC} 
\end{figure}

In Fig. \ref{fig_OP_EC}, to demonstrate the impact of i.n.i.d. fading channels on the system performance, we consider different element settings, i.e., $\vec{L}_2$, $\vec{L}_3$, and $\vec{L}_4$. As can be observed in Fig. \ref{fig_OP_EC}, the number of passive elements installed at the RISs has a significant impact on both the OP and EC. It is because under i.n.i.d. fading channels, each RIS is subject to different fading severity, i.e., the Nakagami-$m$ distributions between different RISs have different shape parameters and spread parameters. When comparing the performance gaps between different element settings, the ORA scheme is more sensitive than the ERA scheme to the changes of element settings. For instance, as shown in Fig.~\ref{fig_OP},  when changing from element setting $\vec{L}_4$ to $\vec{L}_2$, the transmit power of the ERA scheme needs to compensate $1.8$ dBm, whereas the ORA scheme needs to compensate for $6.3$ dBm to achieve the same OP of $10^{-4}$.  In addition, it is evidenced from Fig. \ref{fig_OP_EC} that RIS-aided systems are superior to non-RIS-aided systems, and that the ERA scheme outperforms the ORA scheme. The reason behind this is that the more of the reflecting elements that are installed, the better the performance that can be achieved. For the case of large-scale networks, as in \cite{Lyu_TWC_2021}, we consider that there are two adjacent sources located at $(-99,0)$ and $(255,0)$, respectively, causing co-channel inter-cell interference to the system. As can be observed in Fig.~\ref{fig_OP_EC}, the interference decreases the performance of the considered multi-RIS-aided system.  

\begin{figure}
    \centering
    \begin{minipage}{0.4\textwidth}
        \centering
        \includegraphics[width=.9\textwidth]{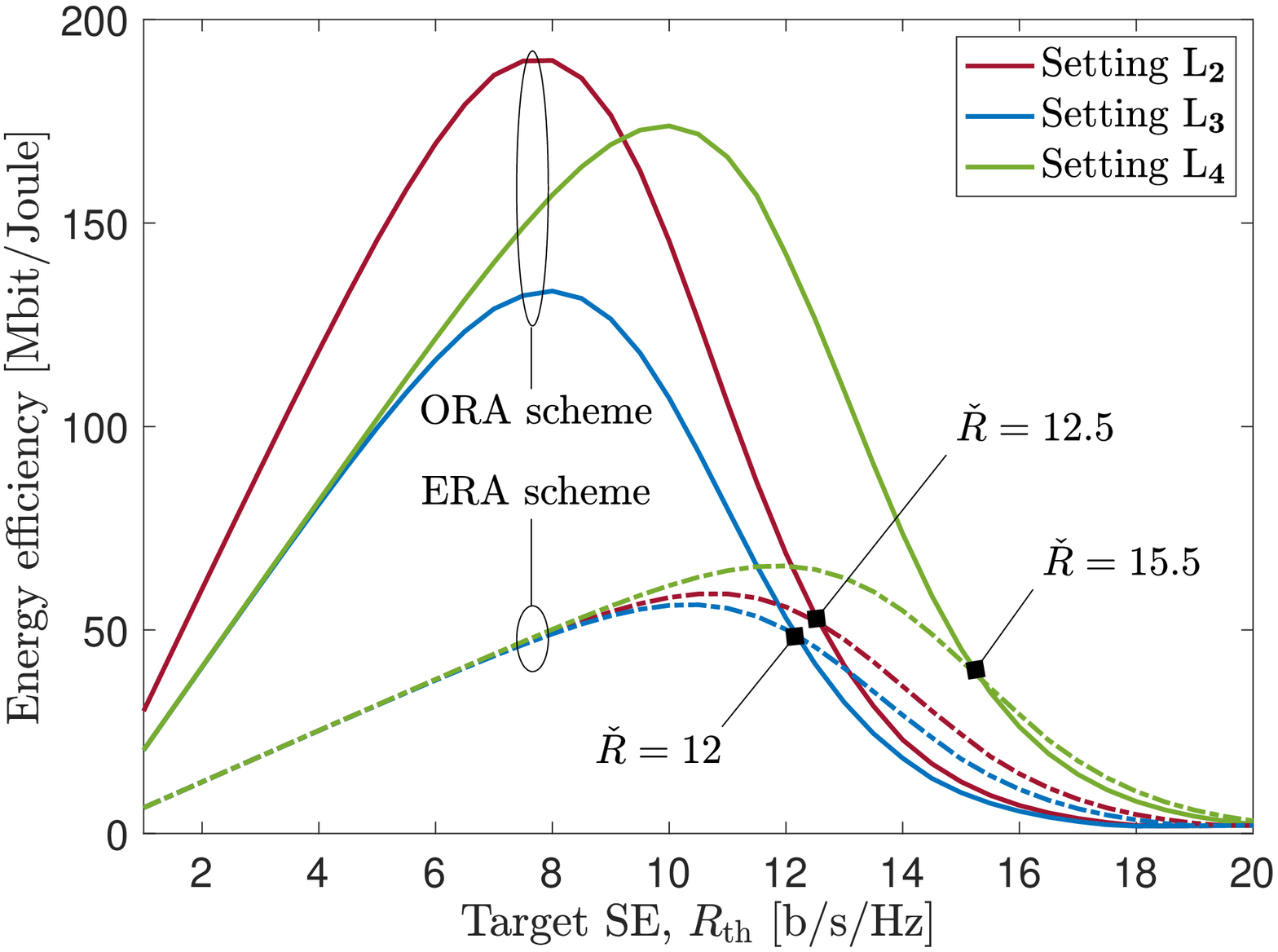} 
        \caption{Energy efficiency [Mbit/Joule] of the ERA and the ORA schemes as a function of target SE, $R_{\rm th}$ [b/s/Hz].}
        \label{fig_EE}
    \end{minipage} \\
    \begin{minipage}{0.4\textwidth}
        \centering
        \includegraphics[width=\textwidth]{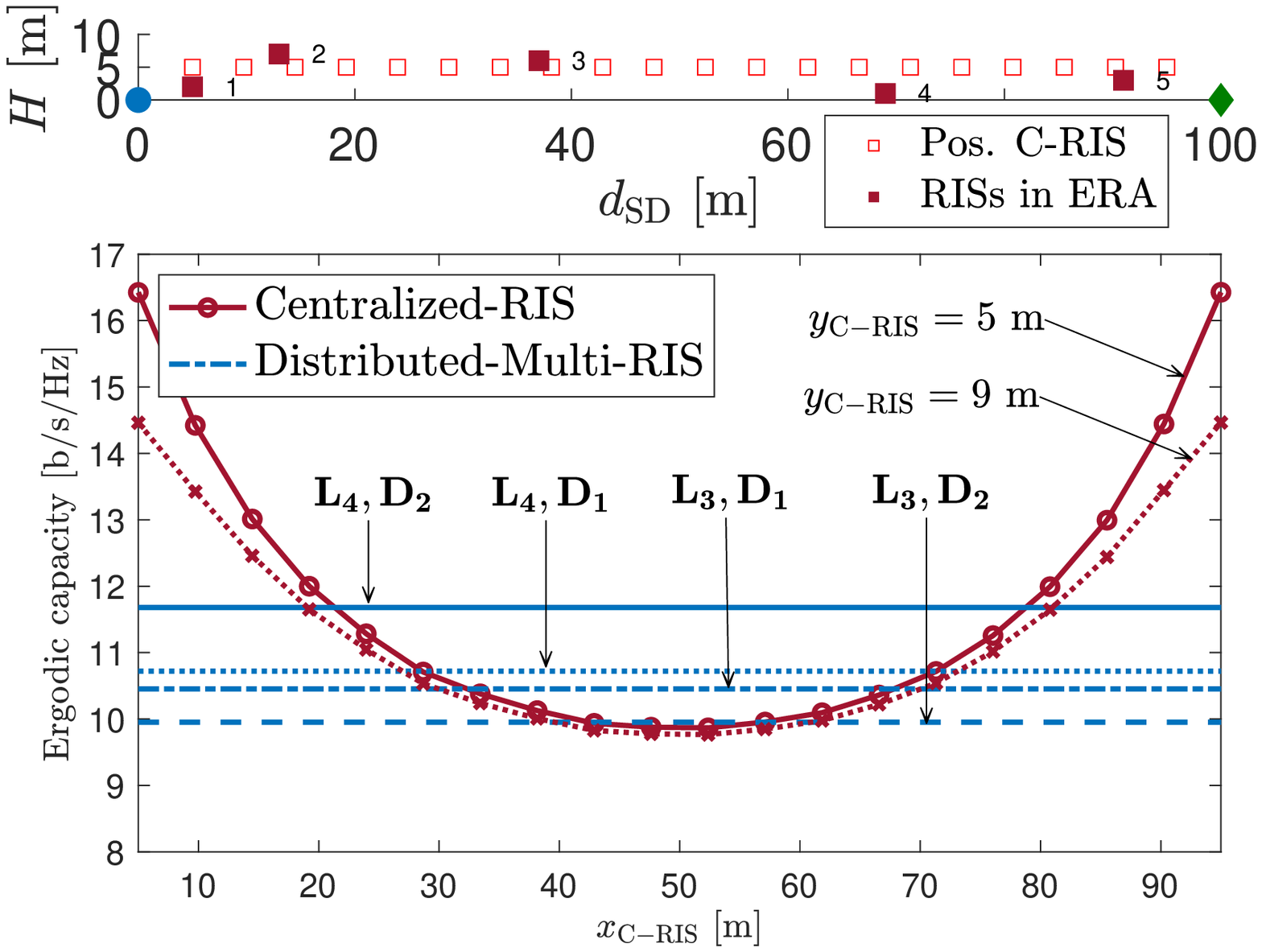} 
        \caption{Performance comparison between centralized versus distributed RIS-aided systems, with $P_\src = 23$ dBm.}
        \label{fig_centralized}
    \end{minipage}
\end{figure}

In Fig. \ref{fig_EE}, taking into account the circuit dissipation power (CDP) in the transceivers and the RIS hardware, we evaluate the EE of the ERA and the ORA schemes. The total power consumed, $P_{\rm tol}$, by a RIS-aided system can be expressed as $P_{\rm tol} = P_\src + \sum_{n=1}^N \sum_{l=1}^{L_n} \tilde{P}_{nl} + \tilde{P}_{\src} + \tilde{P}_\des$ \cite{Huang_TWC_2019}, where $\tilde{P}_{nl}$, $\tilde{P}_{\src}$, $\tilde{P}_\des$ are the CDP at the $l$-th element of the $n$-th RIS, $\src$, and $\des$, respectively. The EE of a RIS-aided system can be defined as $\mathrm{EE} = \mathrm{BW} \times R_{\rm th} / P_{\rm tol}$ [Mbit/Joule] \cite{Bjornson_LWC_2020_b}, where $\mathrm{BW}$ and $R_{\rm th}$ denote the bandwidth and the target SE. As shown in Fig. \ref{fig_EE}, for element setting $\vec{L}_3$,  when $R_{\rm th} \leq \check{R} = 12$ b/s/Hz, where $\check{R}$ denotes the crossing point as illustrated in Fig.~\ref{fig_EE}, the ORA scheme achieves higher EE than the ERA scheme, and vice-versa when $R_{\rm th} \geq \check{R} = 12$ b/s/Hz. In addition, the element setting strongly impacts the EE, i.e., when the element setting is $\vec{L}_2$ and $\vec{L}_4$, the crossing points are  $\check{R} = 12.5$ and $15.5$ b/s/Hz, respectively.

In Fig. \ref{fig_centralized}, we compare the performance of the distributed multi-RIS-aided scheme, i.e., the ERA scheme, versus a centralized RIS (C-RIS)-aided system. Specifically, in the centralized-ERA scheme, we assume that all elements are installed at one large C-RIS, and we let the C-RIS moves along a horizontal line from $\src \to \des$ as depicted in Fig.~\ref{fig_centralized}, where the unfilled red squares are the possible locations of the C-RIS. From various location and reflecting element settings, we can observe a common behavior, i.e., as the C-RIS is located near either $\src$ or $\des$, the centralized-ERA scheme yields a larger EC than the distributed scheme. Meanwhile, when the C-RIS moves further away from either $\src$ or $\des$, the ERA scheme achieves a better EC. \label{page_centralized}

\subsection{The Kullback-Leibler (KL) Divergence and The Kolmogorov-Smirnov (KS) Test}
To gain more insights into the accuracy of the distribution approximation, we carry out the numerical analysis using the Kullback-Leibler (KL) divergence and the Kolmogorov–Smirnov (KS) goodness-of-fit test.
Let $f_X(x)$ and $F_X(x)$ be the true PDF and CDF of an arbitrary random variable $X$. Let $\tilde{f}_X(x)$ and $\tilde{F}_X(x)$ be the approximate PDF and CDF of $X$, respectively.

{\em The Kullback-Leibler (KL) Divergence}:
The KL divergence is a metric that measures the difference between two probability distributions. In our case, the KL divergence of $\tilde{f}_X (x)$ from $f_X (x)$, denoted $\mathrm{D_{KL}} (f_X(x) \Vert \tilde{f}_X (x))$, can be expressed as \cite{Peppas_IET_2011}
\begin{align}
	\mathrm{D_{KL}} (f_X(x) \Vert \tilde{f}_X (x)) &= \int_{-\infty}^{\infty} f_X(x) \ln \frac{f_X(x)}{\tilde{f}_X(x)} dx.
\end{align}
It is noted that the lower the KL divergence value, the better the accuracy of the approximate distribution one can achieve.

{\em The Kolmogorov–Smirnov (KS) Goodness-of-Fit Test}: 
We consider a null hypothesis $H_0$, which states that \textit{the true distribution of the e2e channel coefficient can be approximated by either the Gamma or Log-Normal distributions}; otherwise, it is an alternative hypothesis $H_1$.
The Kolmogorov-Smirnov (KS) goodness-of-fit test relies on the \textit{KS distance}, $\mathrm{D_{KS}}$, which is the maximum value of the absolute difference between the true and approximate distributions, and the \textit{critical value}, $\mathrm{D}_{s,p}$, to decide whether to accept or reject $H_0$. Mathematically, the KS test can be described as \cite{Peppas_IET_2011,KongACCESS2021}
\begin{align}
     \max_s\left\vert F_{X}(\vec{x}) - \tilde{F}_X(\vec{x}) \right\vert \triangleq \mathrm{D_{KS}} \underset{H_0}{\overset{H_1}{\gtrless}} \mathrm{D}_{s,p},
\end{align}
where the critical value $\mathrm{D}_{s,p}$ is given by
%\begin{align}
    $D_{s,p} \approx \sqrt{-(1/2 s )\ln(p/2)}$ \cite{KongACCESS2021},
%\end{align}
where $p$ is the significance level, $s$ is the number of random samples of $X$, and $\vec{x}$ is the $s\times 1$ random sample vector.

\begin{figure}[t]
	\centering
	\subfloat[]{%
		\includegraphics[width=.6\linewidth]{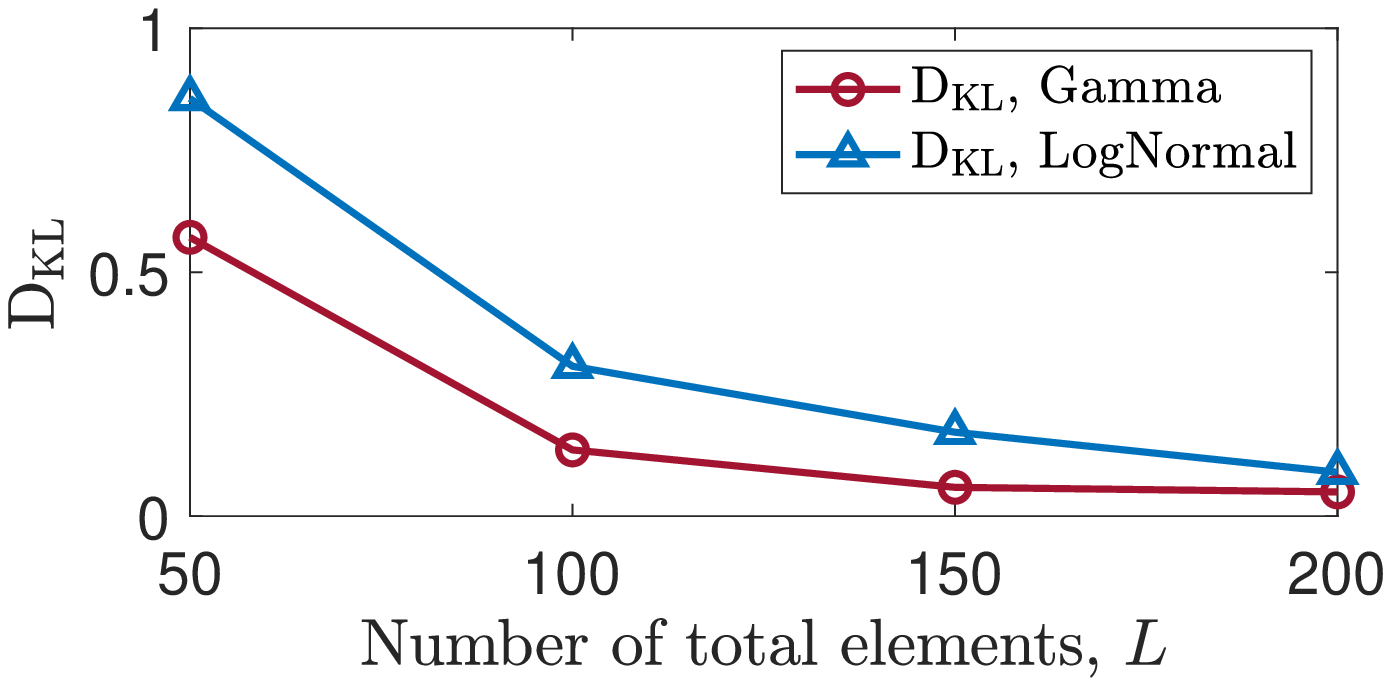}
		\label{fig10_KL}} \\
	\centering
	\subfloat[]{%
		\includegraphics[width=.6\linewidth]{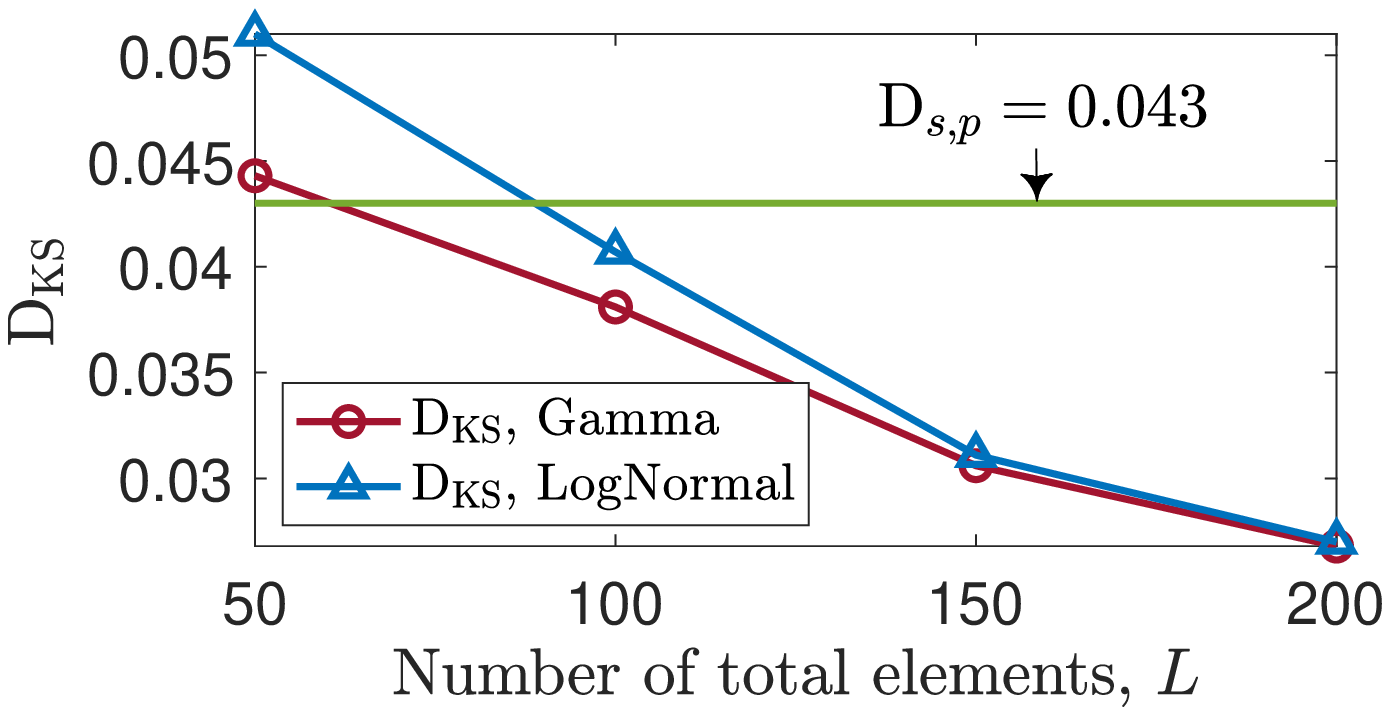}
		\label{fig10_KS}}
	\caption{Demonstration of (a) the Kullback-Leibler divergence, $\mathrm{D_{KL}} (f_X(x) \Vert \tilde{f}_X (x))$, and (b) the Kolmogorov-Smirnov distance, $\mathrm{D_{KS}}$, as a function of the total number of passive reflecting elements, $L = \sum_{n=1}^N L_n$.}
	\label{fig_KL_KS} 
\end{figure}

In Fig.~\ref{fig_KL_KS}, we demonstrate the KL divergence and the KS test, in which, with the help of the Statistics and Machine Learning Toolbox provided by MATLAB \cite{Matlab2021a}, we estimate $f_X(x)$ and $F_X(x)$ using the ${\tt ksdensity}$ and ${\tt ecdf}$ functions, respectively. On one hand, in our KS test, we set $p = 0.05$, $s = 10^3$ samples, and as a result, $D_{s,p} = 0.043$. From Fig.~\ref{fig_KL_KS}, when the total number of the reflecting elements is greater than $100$, the KS test accepts $H_0$ at the significant level of $p = 0.05$. On the other hand, as shown in Fig.~\ref{fig_KL_KS}, when the total number of the reflecting elements is small, the Gamma distribution approximation results in a lower $\mathrm{D_{KL}}$ and $\mathrm{D_{KS}}$ than the Log-Normal distribution approximation, which implies that the Gamma distribution provides a higher accuracy than the Log-Normal distribution. However, when the total number of the reflecting elements is sufficiently large, the Gamma and Log-Normal distributions provide comparable approximation accuracies.

\section{Conclusions} \label{section_conclusions}

In this paper, we proposed two multi-RIS-aided schemes, namely the ERA and the ORA schemes. We focused on the statistical characterization and modeling of the two schemes. Toward this end, we proposed a mathematical framework to determine the distribution of the e2e fading channel in both schemes. Specifically, for the ERA scheme, the framework determined that the true distribution of the magnitude of the e2e channel coefficient can be approximated by either the Gamma or Log-Normal distributions. For the ORA scheme, we obtained approximate closed-form expressions for the CDF and PDF of the magnitude of the e2e channel. Moreover, the framework determined that the ORA scheme's fading channel can also be modeled by a Log-Normal distribution. Based on the resultant fading models, we evaluated the performance of the two schemes in terms of OP and EC. Numerical results showed that the ERA scheme outperforms the ORA scheme in terms of OP and EC. Nevertheless, the ORA scheme achieves better EE than the ERA scheme for some specific values of the target SE. We showed that for a given total number of passive elements, the element allocation and RISs' position have a significant impact on the system performance. On the other hand, the centralized large-RIS-aided system can yield a better EC than the ERA scheme when the centralized RIS is located near either the source or the destination; otherwise, the ERA scheme outperforms the centralized RIS-aided system.

\appendices

\section{Proof of Theorem \ref{theorem_dist_Z_Gamma}} \label{proof_theorem_dist_Z_Gamma}

Since $h_0 \sim \mathrm{Nakagami}(m_0, \Omega_0)$, the $k$-th moment of $h_0$, $\mu_{h_0} (k) \triangleq \mathbb{E}[h_0^k]$, can be obtained as
\begin{align} \label{k_moment_h0}
	\mu_{h_0} (k) = \frac{\Gamma(m_0 + k/2)}{\Gamma(m_0)} \bigg(\frac{m_0}{\Omega_0}\bigg)^{-k/2} .
\end{align}

We now turn our focus to the $k$-th moment of $U_{nl}$. 
The exact PDF of $U_{nl}$ is obtained as
\begin{align} \label{PDF_Unl_exact}
	f_{U_{nl}} (z) &= \frac{ 4 \lambda_{nl}^{m_{\idh_n} + m_{\idg_n}} }{\Gamma(m_{\idh_n}) \Gamma(m_{\idg_n})}  x^{m_{\idh_n} + m_{\idg_n} - 1} K_{m_{\idg_n} - m_{\idh_n}} (2 \lambda_{nl} z),
\end{align}
where $\lambda_{nl} = \sqrt{ \frac{1}{\kappa_{nl}^2} \frac{m_{\idh_n}}{\Omega_{\idh_n}} \frac{m_{\idg_n}}{\Omega_{\idg_n}} }$. 
The detailed derivation of \eqref{PDF_Unl_exact} can be briefly presented as follows. 

The aim of this derivation is not only to present the exact PDF of $U_{nl}$ but also to point out that the magnitude of the synthesized channel of the individual dual-hop channel with respect to a single reflecting element of a RIS in our considered system follows a Generalized-$K$ ($K_G$) distribution.
Recall that, for a given $n$, $U_{nl} = \kappa_{nl} h_{nl}g_{nl}$, where $U_{n1},\ldots,U_{nL_n}$ are i.i.d. RVs, but $h_{nl}$ and $g_{nl}$ are i.n.i.d. RVs, $\forall n, \forall l$. Specifically, knowing that $f_{XY} (z) = \int_0^\infty \frac{1}{x} f_Y \left(\frac{z}{x}\right) f_X(x) dx$, the PDF of $U_{nl}$ can be written as
\begin{equation}
	f_{U_{nl}} (z)  = \frac{1}{\kappa_{nl}} \int_{0}^\infty \frac{1}{x} f_{h_{nl}} \left( \frac{z}{ \kappa_{nl} x} \right) f_{g_{nl}} (x) dx.
\end{equation}
Recall that  $h_{nl} \sim \mathrm{Nakagami}(m_{\idh_n}, \Omega_{\idh_n})$ and $g_{nl} \sim \mathrm{Nakagami}(m_{\idg_n}, \Omega_{\idg_n})$, from \eqref{PDF_Naka}, we have
\begin{align}
	&f_{U_{nl}} (z) = \frac{4}{\Gamma(m_{\idh_n}) \Gamma(m_{\idg_n})} \left(\frac{1}{\kappa_{nl}^2} \!\!\!\!\! \frac{m_{\idh_n}}{\Omega_{\idh_n}}\right)^{m_{\idh_n}} \!\!\! \left(\frac{m_{\idg_n}}{\Omega_{\idg_n}}\right)^{m_{\idg_n}} \!\!\!\!
	z^{2 m_{\idh_n} - 1} \nonumber \\
	&\quad\times \int_{0}^{\infty} x^{2 m_{\idg_n} - 2 m_{\idh_n} - 1}
	e^{ \left( - \frac{z^2}{\kappa_{nl}^2} \frac{m_{\idh_n}}{\Omega_{\idh_n}} \frac{1}{x^2} - \frac{m_{\idg_n}}{\Omega_{\idg_n}} x^2 \right)} dx.
\end{align}
Making use of \cite[Eq. (3.478.4)]{Gradshteyn2007}, the exact PDF of $U_{nl}$ is obtained as in \eqref{PDF_Unl_exact}. Thus, we can conclude that $U_{nl}$ follows a $K_G$ distribution with the shaping parameters $m_{\idg_n}$ and $m_{\idh_n}$.
Based on the exact PDF of $U_{nl}$, we are going to derive the $k$-th moment of $U_{nl}$, which is defined as $\mu_{U_{nl}} (k) \triangleq \mathbb{E}[U_{nl}^k] = \int_0^\infty z^k f_{U_{nl}} (z) dz$. From \eqref{PDF_Unl_exact} and after some mathematical manipulations, $\mu_{U_{nl}} (k)$ can be obtained as
\begin{align} 
	\mu_{U_{nl}} (k) 
	=
	\lambda_{nl}^{-k} \frac{ \Gamma(m_{\idh_n} + k/2) \Gamma(m_{\idg_n} + k/2) }{ \Gamma(m_{\idh_n}) \Gamma(m_{\idg_n}) }.
	\label{eq_k_th_moment_U_nl}
\end{align}
The exact PDF of $U_{nl}$ in \eqref{PDF_Unl_exact} helps derive the $k$-th moment of $U_{nl}$, however, it makes the derivation of closed-form expressions for the CDF and PDF of $Z$ intractable. To circumvent this problem, based on the obtained $k$-th moment of $U_{nl}$ in \eqref{eq_k_th_moment_U_nl}, we fit $U_{nl}$ to a $\rm Gamma$ distribution. Note that for a given $n$, the $U_{nl}$ RVs are i.i.d. with respect to $l = 1, \ldots, L_n$. Relying on Step 2 in our framework, we have
\begin{align}
	U_{nl} \overset{\rm approx.}{\sim} \mathrm{Gamma} (\alpha_{\idu_n},\beta_{\idu_n}) ,
\end{align} 
where
\begin{align}
	\alpha_{\idu_n} = \frac{ \left(\mathbb{E}[U_{nl}]\right)^2 }{\mathrm{Var}[U_{nl}]} = \frac{[\mu_{U_{nl}}(1)]^2}{\mu_{U_{nl}}(2) - [\mu_{U_{nl}}(1)]^2}, \\
	\beta_{\idu_n} = \frac{ \mathbb{E}[U_{nl}] }{\mathrm{Var}[U_{nl}]} = \frac{\mu_{U_{nl}}(1)}{\mu_{U_{nl}}(2) - [\mu_{U_{nl}}(1)]^2} ,
\end{align}
and $f_{U_{nl}} (z; \alpha_{\idu_n}, \beta_{\idu_n})$ is expressed as in \eqref{PDF_Gamma}.
The analysis will rely on the approximate PDF of $U_{nl}$. In addition, for a given $n$-th RIS, we consider fixed amplitude reflection coefficients \cite{Bjornson_WCL_2020}, i.e., $\kappa_{nl} = \kappa_n, \forall l$. 

The approximate distribution of $V_n$ can be obtained as
\begin{align} \label{eq_Gamma_V_n}
	V_n \overset{\rm approx.}{\sim} \mathrm{Gamma}(L_n \alpha_{\idu_n}, \beta_{\idu_n}).
\end{align}
Consequently, the approximate CDF and PDF of $V_n$ are obtained as
\begin{align}
	F_{V_n} (z) &\approx \frac{ \gamma(L_n \alpha_{\idu_n}, \beta_{\idu_n} z) }{ \Gamma \left(L_n \alpha_{\idu_n} \right) }, \label{CDF_Vn_end} \\
	f_{V_n} (z) &\approx f_{V_n}(z; L_n \alpha_{\idu_n}, \beta_{\idu_n}), \label{PDF_Vn_end}
\end{align}
where $f_{V_n}(z; L_n \alpha_{\idu_n}, \beta_{\idu_n})$ is expressed as in \eqref{PDF_Gamma}.

Making use of a multinomial expansion \cite{Costa_CL_2011}, the $k$-th moment of $V_n$, i.e., $\mu_{V_n} (k) \triangleq \mathbb{E}[V_n^k]$, can be obtained as
\begin{align} \label{mu_Vn_k}
	\mu_{V_n} (k) 
	&= \sum_{k_1 = 0}^{k} \sum_{k_2 = 0}^{k_1} \cdots \sum_{k_{L_n - 1} =0}^{k_{L_n - 2}} 
		\binom{k}{k_1} \binom{k_1}{k_2} \cdots \binom{k_{L_n - 2}}{k_{L_n -1}} \nonumber \\
	&\quad \times \mu_{U_{n1}} (k-k_1) \mu_{U_{n2}} (k_1-k_2)\ldots \mu_{U_{nL_n}} (k_{L_n - 1}).
\end{align}

We turn our focus to the $k$-th moment of $T$, i.e., $\mu_T (k) \triangleq \mathbb{E}[T^k]$. Proceeding in a similar manner, the $k$-th moment of $T$, i.e., $\mu_T (k) \triangleq \mathbb{E}[T^k]$, can be obtained as
\begin{align} \label{mu_T_k}
	\mu_T (k)	&= \sum_{k_1 = 0}^{k} \sum_{k_2 = 0}^{k_1} \cdots \sum_{k_{N-1} =0}^{k_{N-2}} 
			\binom{k}{k_1} \binom{k_1}{k_2} \cdots \binom{k_{N-2}}{k_{N-1}} \nonumber \\
		&\quad \times \mu_{V_1} (k-k_1) \mu_{V_2} (k_1-k_2)\ldots \mu_{V_n} (k_{N - 1}).
\end{align}
From \eqref{eq_k_th_moment_U_nl}, \eqref{mu_Vn_k}, and \eqref{mu_T_k}, the first and second moments of $T$ can be expressed as
\begin{align} 
    \mu_T (1) &= \sum_{n=1}^{N} \sum_{l=1}^{L_n} \mu_{U_{nl}}(1), \label{mu_T_1} \\
    \mu_T (2) &= \sum_{n=1}^{N} \bigg[\sum_{l=1}^{L_n} \mu_{U_{nl}} (2) + 2 \sum_{l=1}^{L_n} \mu_{U_{nl}} (1) \sum_{l'=l+1}^{L_n} \mu_{U_{nl'}}(1) \bigg] \nonumber \\
    &\quad+ 2\sum_{n=1}^N \bigg[\sum_{l=1}^{L_n}\mu_{U_{nl}}(1)\bigg] \sum_{n'=n+1}^{N} \bigg[ \sum_{l=1}^{L_{n'}} \mu_{U_{n'l}}(1)\bigg].  \label{mu_T_2}
\end{align}

Since $h_0$ and $T$ are independent, the $k$-th moment of $Z$, $\mu_Z (k) \triangleq \mathbb{E}[Z^k]$, can be obtained via the moments of its summands, i.e., $h_0$ and $T$, by applying the binomial theorem. Thus, $\mu_Z (k)$ can be obtained as
\begin{align} \label{mu_Z_k}
	\mu_Z (k) &= \mathbb{E}[(h_0 + T)^k] = \mathbb{E}\bigg[\sum_{l=0}^k \binom{k}{l} h_0^l T^{k-l}\bigg] \nonumber \\
	&= \sum_{l=0}^k \binom{k}{l} \mu_{h_0}(l) \mu_T (k-l) .
\end{align}
Thus, from \eqref{mu_Z_k}, we have
\begin{align}
	\mu_Z (1) &= \mu_{h_0} (1) + \mu_T (1), \label{mu_Z_1} \\
	\mu_Z (2) &= \mu_{h_0} (2) + \mu_T (2) + 2 \mu_{h_0} (1) \mu_T (1) . \label{mu_Z_2}
\end{align}
From \eqref{k_moment_h0}, we have 
\begin{align}
	\mu_{h_0} (1) &= \frac{\Gamma(m_0 + 1/2)}{\Gamma(m_0)} \bigg(\frac{m_0}{\Omega_0}\bigg)^{-1/2}, \label{mu_h0_1} \\
	\mu_{h_0} (2) &= \frac{\Gamma(m_0 + 1)}{\Gamma(m_0)}\frac{\Omega_0}{m_0} = \Omega_0 . \label{mu_h0_2}
\end{align}
Plugging \eqref{mu_h0_1} and \eqref{mu_T_1} into \eqref{mu_Z_1}, and plugging \eqref{mu_h0_2} and \eqref{mu_T_2} into \eqref{mu_Z_2}, we complete the proof of Theorem \ref{theorem_dist_Z_Gamma}.

\section{Derivation of \eqref{eq_ERA_Gamma_EC_end}} \label{appx_ec_ctt_Gamma}

From \eqref{eq_ERA_Gamma_EC_end}, the EC of the ERA scheme can be rewritten as
\begin{align}
	\bar{C}^{\era,\rm Gam}
	&= \int_{0}^\infty \log_2 \left( 1 + z \right) f_{\avgSNR Z^2} (z) dz \nonumber \\
	&= \frac{1}{\ln 2} \int_{0}^\infty \frac{1}{z+1} \left[ 1 - F_Z \left( \sqrt{\frac{z}{\avgSNR}} \right) \right] dz .  
	\label{ec_cct_gamma_a_b}
\end{align}
Making use of the identity \cite[Eq. (8.4.2.5)]{Prudnikov1999}, i.e., $(1+x)^{-\xi} = \frac{1}{\Gamma(\xi)} G_{1,1}^{1,1} \left( x \middle\vert
\begin{matrix}
1-\xi \\ 0
\end{matrix} \right)$, \eqref{ec_cct_gamma_a_b} can be further expressed as
\begin{align}
	\bar{C}^{\era, \rm Gam} \approx \frac{1}{\ln 2} \int_{0}^\infty 
	G_{1,1}^{1,1} \left( z \middle\vert
	\begin{matrix}
	0 \\ 0
	\end{matrix} \right)
	\frac{ \Gamma \left( {\alpha_\idz} , {\beta_\idz}  \sqrt{\frac{z}{\avgSNR}}  \right) }{ \Gamma({\alpha_\idz})} dz. \label{ec_cct_gamma_b}
\end{align}
Based on \cite[Eq. (8.356.3)]{Gradshteyn2007}, i.e., $\Gamma(\alpha,x) + \gamma(\alpha,x) = \Gamma(\alpha)$, we have
\begin{align}
	F_{Z} (z) = \frac{ \gamma \left( {\alpha_\idz} , {\beta_\idz} \sqrt{\frac{z}{\avgSNR}}  \right) }{ \Gamma({\alpha_\idz})} = 1 - \frac{ \Gamma \left( {\alpha_\idz} , {\beta_\idz} \sqrt{\frac{z}{\avgSNR}}  \right) }{ \Gamma({\alpha_\idz})}.
\end{align}
Using the equivalence between the incomplete Gamma function and the Meijer-G function \cite[Eq. (8.4.16.2)]{Prudnikov1999}, i.e., $\Gamma \left( {\alpha_\idz} , {\beta_\idz} \sqrt{\frac{z}{\avgSNR}}  \right) = G^{2,0}_{1,2} \left( {\beta_\idz} \sqrt{\frac{z}{\avgSNR}} \middle\vert
	\begin{matrix}
	1 \\ {\alpha_\idz}, 0
	\end{matrix} \right)$, \eqref{ec_cct_gamma_b} can be further expressed as
\begin{align} \label{eq_EC_Meijer}
	&\bar{C}^{\era, \rm Gam} \nonumber \\
	&\approx \frac{1}{\Gamma({\alpha_\idz}) \ln 2 } \int_{0}^\infty 
		G_{1,1}^{1,1} \left( z \middle\vert
		\begin{matrix}
		0 \\ 0
		\end{matrix} \right)
		G^{2,0}_{1,2} \left( {\beta_\idz} \sqrt{\frac{z}{\avgSNR}} \middle\vert
			\begin{matrix}
			1 \\ {\alpha_\idz},0
			\end{matrix} \right) dz.
\end{align}

Next, we rely on the identity \cite[Eq. (2.24.1.1)]{Prudnikov1999} for the general integral of 
\begin{align*}
\int_{0}^{\infty} x^{\alpha-1} G^{s,t}_{u,v} \left( \xi x \middle\vert
			\begin{matrix}
			(c_u) \\ (d_v)
			\end{matrix} \right)
			G^{m,n}_{p,q} \left( \omega x^{l/k} \middle\vert
						\begin{matrix}
						(a_p) \\ (b_q)
						\end{matrix} \right) dx .
\end{align*}
Thus, the EC of the ERA scheme can be obtained as
\begin{align}
	&\bar{C}^{\era, \rm Gam} \nonumber \\
	&\approx \frac{1}{\Gamma({\alpha_\idz}) \ln 2 } \frac{2^{{\alpha_\idz} -1}}{\sqrt{\pi}}
	G^{5,1}_{3,5} \left( \frac{ ({\beta_\idz})^2 }{4  \avgSNR} \middle\vert
					\begin{matrix}
					0, \frac{1}{2}, 1 \\ \frac{{\alpha_\idz}}{2}, \frac{{\alpha_\idz} +1}{2},0,\frac{1}{2},0
					\end{matrix} \right) .
\end{align}

\section{Proof of Lemma \ref{lemma_k_th_moment_M_V}} \label{proof_lemma_k_th_moment_M_V}

The $k$-moment of $M_V$, $\mu_{M_V} (k) \triangleq \mathbb{E} [M_V^k]$, can be expressed as
\begin{align}
	\mu_{M_V} (k) = \sum_{n=1}^N \int_0^\infty x^k f_{V_n} (x) \prod_{ \substack{t=1 \\ t \neq n}}^N F_{V_t} (x) dx . \label{mu_MV_b}
\end{align}
Thus, we have
\begin{align}
	\mu_{M_V} (k) 
	\approx
	\sum_{n=1}^N
	\int_{0}^\infty x^k f_{V_n} (x) 
	\prod_{ \substack{t=1 \\ t \neq n}}^N \frac{ \gamma(L_t {\alpha_\idu}_t , {\beta_\idu}_t) }{ \Gamma(L_t {\alpha_\idu}_t) } dx .
\end{align}
The $\gamma(\cdot,\cdot)$ in \eqref{CDF_Vn_end} can be re-expressed as \cite[Eq. (6.5.29)]{Abramowitz1972}
\begin{align}
	\frac{ \gamma(L_t \alpha_{\idu_n}, {\beta_\idu}_t x) }{ \Gamma(L_t {\alpha_\idu}_t) } 
	= 
	e^{-{\beta_\idu}_t x} \sum_{w_t=0}^{\infty} 
	\frac{ ({\beta_\idu}_t x)^{L_t {\alpha_\idu}_t + w_t} }{ \Gamma( L_t {\alpha_\idu}_t + w_t + 1 ) } . \label{incomplete_gamma_expansion_b}
\end{align}
Plugging \eqref{incomplete_gamma_expansion_b} into $\prod_{ \substack{t=1 \\ t \neq n}}^N F_{{V}_t} (x)$, after some mathematical manipulations, we have
\begin{align}
	&\prod_{ \substack{t=1 \\ t \neq n}}^N \frac{ \gamma(L_t {\alpha_\idu}_t , {\beta_\idu}_t) }{ \Gamma(L_t {\alpha_\idu}_t) } \nonumber \\
	&\quad= 
	e^{- x \sum_{\substack{t=1 \\ t \neq n}}^N {\beta_\idu}_t}
	\widetilde{\sum_{\substack{w_t \\ t \neq n}}}	\bigg[ \prod_{\substack{t=1 \\ t \neq n}}^N \frac{ ({\beta_\idu}_t x)^{L_t {\alpha_\idu}_t + w_t} }{ \Gamma( L_t {\alpha_\idu}_t + w_t + 1 ) } \bigg] ,
\end{align}
where
\begin{align*}
\widetilde{\sum_{\substack{w_t \\ t \neq n}}} \triangleq \sum_{w_1=0}^\infty \ldots \sum_{w_{n-1}=0}^\infty \sum_{w_{n+1}=0}^\infty \ldots \sum_{w_N=0}^\infty.
\end{align*}
Thus, we have
\begin{align}
	&\mu_{M_V} (k) 
	\approx 
	\sum_{n=1}^N
	\widetilde{\sum_{\substack{w_t \\ t \neq n}}}	
	\bigg[ 
	\prod_{\substack{t=1 \\ t \neq n}}^N 
	\frac{ ({\beta_\idu}_t x)^{L_t {\alpha_\idu}_t + w_t} }
	{ \Gamma( L_t {\alpha_\idu}_t + w_t + 1 ) } 
	\bigg]
	\nonumber \\
	&\times
	\int_0^\infty 
	\frac{ (\beta_{\idu_n})^{L_n \alpha_{\idu_n}} }
	{ \Gamma(L_n \alpha_{\idu_n})} 
	x^{ \sum_{j=1}^N L_j {\alpha_\idu}_j 
	+ 
	\sum_{\substack{t=1 \\ t \neq n}} w_t + k -1  }
	e^{-x \sum_{i=1}^N {\beta_\idu}_i } dx .
\end{align}
Let us denote $\Psi_{\check{n}} \triangleq \sum_{\substack{t=1 \\ t \neq n}}^N w_t$ and $\Lambda \triangleq \sum_{t=1}^N L_t {\alpha_\idu}_t$. After some mathematical manipulations, $\mu_{M_V}(k)$ can be expressed as
 \begin{align}
 \mu_{M_V} (k) 
 	&\approx 
 	\sum_{n=1}^N
 	\widetilde{\sum_{\substack{w_t \\ t \neq n}}}	\bigg[ \prod_{\substack{t=1 \\ t \neq n}}^N \frac{ ({\beta_\idu}_t x)^{L_t {\alpha_\idu}_t + w_t} }{ \Gamma( L_t {\alpha_\idu}_t + w_t + 1 ) } \bigg]
 	\frac{ (\beta_{\idu_n})^{L_n \alpha_{\idu_n}}}{ \Gamma(L_n \alpha_{\idu_n}) }
 	\nonumber\\
 	&\quad\times
 	\int_0^\infty 
 	x^{\Lambda + \Psi_{\check{n}} + k - 1} 
 	e^{-x\sum_{i=1}^N {\beta_\idu}_i} dx . \label{mu_MV_c}
 \end{align}
Making use of the identity $\int_0^\infty x^{v-1} e^{-px} dx = \Gamma(v) / p^v$ \cite[Eq. (3.381.4)]{Gradshteyn2007}, the integral in \eqref{mu_MV_c} can be derived as
\begin{align}
 \mu_{M_V} (k) 
 	&\approx 
 	\sum_{n=1}^N
 	\bigg\{
 	\widetilde{\sum_{\substack{w_t \\ t \neq n}}}	
 	\bigg[ 
 	\prod_{\substack{t=1 \\ t \neq n}}^N 
 	\frac{ ({\beta_\idu}_t x)^{L_t {\alpha_\idu}_t + w_t} } \nonumber\\
 	&\times
 	{ \Gamma( L_t {\alpha_\idu}_t + w_t + 1 ) } 
 	\bigg]
 	\frac{ (\beta_{\idu_n})^{L_n \alpha_{\idu_n}}}{ \Gamma(L_n \alpha_{\idu_n}) }
 	\Gamma(\Lambda + \Psi_{\check{n}} + k) \nonumber\\
 	&\times
 	\bigg[ \sum_{i=1}^N {\beta_\idu}_i \bigg]^{- (\Lambda + \Psi_{\check{n}} + k)}
 	\bigg\}
 	.
\end{align}
After some rearrangements, we have
%
%\begin{figure*}
\begin{align}
	\mu_{M_V} (k)
	 	&\approx 
	 	\sum_{n=1}^N
	 	\bigg\{
	 	\frac{1}{ \Gamma(L_n {\alpha_\idu}_t) }
	 	\bigg[ 
	 	(\beta_{\idu_n})^{L_n \alpha_{\idu_n}} 
	 	\prod_{\substack{t=1 \\ t \neq n}}^N ({\beta_\idu}_t)^{L_t {\alpha_\idu}_t} 
	 	\bigg] \nonumber \\
	 	&\times 
	 	\bigg[ \sum_{t=i}^N {\beta_\idu}_i \bigg]^{-(\Lambda + k)}
	 	\widetilde{\sum_{\substack{w_t \\ t \neq n}}}
	 	\Gamma(\Lambda + \Psi_{\check{n}} + k)	\nonumber \\
	 	&\times	 	
	 	\bigg[ \prod_{\substack{t=1 \\ t \neq n}}^N \frac{ ({\beta_\idu}_t)^{w_t} }{ \Gamma( L_t {\alpha_\idu}_t + w_t + 1 ) } \bigg]
	 	\bigg[ \sum_{i=1}^N {\beta_\idu}_i \bigg]^{-\Psi_{\check{n}}}
	 	\bigg\} . \label{temp}
\end{align}
%\end{figure*} 

Let $\chi_t = {\beta_\idu}_t / \sum_{v=1}^N {\beta_\idu}_v$. It can be observed that $\chi_t \in (0,1)$ and $\sum_{t=1}^N \chi_t = 1$, which yields
\begin{align}
	\bigg[ (\beta_{\idu_n})^{L_n \alpha_{\idu_n}} \prod_{\substack{t=1 \\ t \neq n}}^N {\beta_\idu}_t^{L_t {\alpha_\idu}_t} \bigg] 
	\bigg[ \sum_{i=1}^N {\beta_\idu}_i \bigg]^{(-\Lambda)}
	&=
	\prod_{j=1}^N (\chi_j)^{L_j {\alpha_\idu}_j}, \label{temp1}	
\end{align}
By plugging \eqref{temp1} into \eqref{temp}, we have
\begin{align}
	\mu_{M_V} (k)
	 	&\approx 
	 	\sum_{n=1}^N
	 	\bigg\{
	 	\frac{1}{ \Gamma(L_n {\alpha_\idu}_t) }
	 	\bigg[ \prod_{j=1}^N (\chi_j)^{L_j {\alpha_\idu}_j} \bigg] \nonumber \\
	 	&\times
	 	\bigg[ \sum_{i=1}^N {\beta_\idu}_i \bigg]^{-k}
	 	\widetilde{\sum_{\substack{w_t \\ t \neq n}}}
	 	\Gamma(\Lambda + \Psi_{\check{n}} + k)	\nonumber \\
	 	&\times	 	
	 	\bigg[ \prod_{\substack{t=1 \\ t \neq n}}^N \frac{ ({\beta_\idu}_t)^{w_t} }{ \Gamma( L_t {\alpha_\idu}_t + w_t + 1 ) } \bigg]
	 	\bigg[ \sum_{i=1}^N {\beta_\idu}_i \bigg]^{-\Psi_{\check{n}}}
	 	\bigg\} . \label{temp3}
\end{align} 
After some mathematical manipulations, it follows that 
\begin{align}
	&\bigg[ 
	\prod_{\substack{t=1 \\ t \neq n}}^N 
	\frac{ ({\beta_\idu}_t)^{w_t} }{ \Gamma( L_t {\alpha_\idu}_t + w_t + 1 ) } 
	\bigg]
	\bigg[ \sum_{i=1}^N {\beta_\idu}_i \bigg]^{-\Psi_{\check{n}}} \nonumber \\
	&\quad =
	\prod_{ \substack{t=1 \\ t \neq n}}
	\frac{1}
	{\Gamma (L_t {\alpha_\idu}_t + w_t +1) }
	\prod_{ \substack{t=1 \\ t \neq n}} 
	(\chi_t)^{w_t} . \label{temp4}
\end{align}
Note that, in \eqref{temp4}, $i$ is independent from $t$. Next, let $\langle x \rangle_n \triangleq \Gamma(x+n)/\Gamma(x)$ denote a \textit{Pochhammer symbol}, which implies
\begin{align}
	\Gamma(\Lambda + k + \Psi_{\check{n}}) = \langle \Lambda+k \rangle_{\Psi_{\check{n}}} \Gamma(\Lambda + k) .
\end{align}
Thus, the right-hand-side (RHS) of \eqref{temp4} can be expressed as
\begin{align}
	&\prod_{ \substack{t=1 \\ t \neq n}}
	\frac{ (\chi_t)^{w_t} }
	{\Gamma (L_t {\alpha_\idu}_t + w_t +1) }\nonumber \\
	&=
	\bigg[ 
	\prod_{ \substack{t=1 \\ t \neq n}}
	\frac{1}{\Gamma (L_t {\alpha_\idu}_t + 1)} 
	\bigg]
	\bigg[
	\prod_{ \substack{t=1 \\ t \neq n}}
	\frac{ \langle 1 \rangle_{w_t} }{ \langle L_t {\alpha_\idu}_t + 1\rangle_{w_t}}
	\bigg]
	\bigg[
	\prod_{ \substack{t=1 \\ t \neq n}}^N 
	\frac{ (\chi_t)^{w_t} }{w_t !}
	\bigg] . \label{temp5}
\end{align}
After some mathematical manipulations, we have
\begin{align}
	 & \mu_{M_V} (k)
	\approx
	\bigg[ \sum_{i=1}^N {\beta_\idu}_i \bigg]^{-k}
	\bigg[ \prod_{j=1}^N (\chi_j)^{L_j {\alpha_\idu}_j} \bigg] 
	\nonumber \\
	 & \times
	\sum_{n=1}^N 
	\bigg\{
	\frac{ \Gamma(\Lambda+k) }{ \Gamma(L_n \alpha_{\idu_n}) } 
	\bigg[
	\prod_{ \substack{t=1 \\ t \neq n}}^N \frac{1}{ \Gamma(L_n \alpha_{\idu_n} + 1) }
	\bigg] \nonumber \\
	&\times
	\widetilde{\sum_{\substack{w_t \\ t \neq n}}}
	\bigg[
	\frac{
	\langle \Lambda+k \rangle_{ \sum_{\substack{t=1 \\ t \neq n}}^N w_t } 
	\prod_{ \substack{t=1 \\ t \neq n}}^N \langle 1 \rangle_{w_t} 
	}
	{
	\prod_{ \substack{t=1 \\ t \neq n}}^N \langle L_t {\alpha_\idu}_t + 1 \rangle_{w_t}  
	}
	\prod_{ \substack{t=1 \\ t \neq n}}^N \frac{ (\chi_t)^{w_t} }{w_t !}
	\bigg]
	\bigg\} .
\end{align}
Using the series representation of the Lauricell function Type-A \cite[Eq.(1.4.1)]{Srivastava1985}, and relying on its integral representation \cite[Eq.(9.4.35)]{Srivastava1985}, we obtain the approximate closed-form expression for $\mu_{M_V} (k)$ in \eqref{mu_M_V_k_end}. This completes the proof of Lemma \ref{lemma_k_th_moment_M_V}. 

\balance
\bibliographystyle{IEEEtran}
\bibliography{Refs_Multi_RIS}

\begin{thebibliography}{10}
\providecommand{\url}[1]{#1}
\csname url@samestyle\endcsname
\providecommand{\newblock}{\relax}
\providecommand{\bibinfo}[2]{#2}
\providecommand{\BIBentrySTDinterwordspacing}{\spaceskip=0pt\relax}
\providecommand{\BIBentryALTinterwordstretchfactor}{4}
\providecommand{\BIBentryALTinterwordspacing}{\spaceskip=\fontdimen2\font plus
\BIBentryALTinterwordstretchfactor\fontdimen3\font minus
  \fontdimen4\font\relax}
\providecommand{\BIBforeignlanguage}[2]{{%
\expandafter\ifx\csname l@#1\endcsname\relax
\typeout{** WARNING: IEEEtran.bst: No hyphenation pattern has been}%
\typeout{** loaded for the language `#1'. Using the pattern for}%
\typeout{** the default language instead.}%
\else
\language=\csname l@#1\endcsname
\fi
#2}}
\providecommand{\BIBdecl}{\relax}
\BIBdecl

\bibitem{DiRenzo_JSAC_2020}
M.~{Di Renzo}, A.~Zappone, M.~Debbah, M.-S. Alouini, C.~Yuen, J.~de~Rosny, and
  S.~Tretyakov, ``Smart radio environments empowered by reconfigurable
  intelligent surfaces: How it works, state of research, and the road ahead,''
  \emph{IEEE J. Sel. Areas Commun.}, vol.~38, no.~11, pp. 2450--2525, Nov.
  2020.

\bibitem{Wu_TWC_2019}
Q.~Wu and R.~Zhang, ``Intelligent reflecting surface enhanced wireless network
  via joint active and passive beamforming,'' \emph{{IEEE} Trans. Wireless
  Commun.}, vol.~18, no.~11, pp. 5394--5409, Nov. 2019.

\bibitem{Liaskos_TCOM_2020}
C.~Liaskos, S.~Nie, A.~Tsioliaridou, A.~Pitsillides, S.~Ioannidis, and
  I.~Akyildiz, ``End-to-end wireless path deployment with intelligent surfaces
  using interpretable neural networks,'' \emph{IEEE Trans. Commun.}, vol.~68,
  no.~11, pp. 6792--6806, Nov. 2020.

\bibitem{Galappaththige_arXiv_2020}
\BIBentryALTinterwordspacing
D.~L. Galappaththige, D.~Kudathanthirige, and G.~A.~A. Baduge, ``Performance
  analysis of distributed intelligent reflective surfaces for wireless
  communications,'' Oct. 2020. [Online]. Available:
  \url{http://arxiv.org/abs/2010.12543}
\BIBentrySTDinterwordspacing

\bibitem{Tahir_LWC_2020}
B.~Tahir, S.~Schwarz, and M.~Rupp, ``Analysis of uplink {IRS}-assisted {NOMA}
  under {Nakagami-m} fading via moments matching,'' \emph{IEEE Wireless Commun.
  Lett.}, 2020, {DOI: 10.1109/LWC.2020.3043810}.

\bibitem{Basar_ACCESS_2019}
E.~Basar, M.~{Di Renzo}, J.~{De Rosny}, M.~Debbah, M.-S. Alouini, and R.~Zhang,
  ``Wireless communications through reconfigurable intelligent surfaces,''
  \emph{IEEE Access}, vol.~7, pp. 116\,753--116\,773, Jun. 2019.

\bibitem{Figueiredo_ACCESS_2021}
F.~A.~P. de~Figueiredo, M.~S.~P. Facina, R.~C. Ferreira, Y.~Ai, R.~Ruby, Q.-V.
  Pham, and G.~Fraidenraich, ``Large intelligent surfaces with discrete set of
  phase-shifts communicating through double-rayleigh fading channels,''
  \emph{IEEE Access}, vol.~9, pp. 20\,768--20\,787, Feb. 2021.

\bibitem{Badiu_LWC_2020}
M.-A. Badiu and J.~P. Coon, ``Communication through a large reflecting surface
  with phase errors,'' \emph{IEEE Wireless Commun. Lett.}, vol.~9, no.~2, pp.
  184--188, Feb. 2020.

\bibitem{Boulogeorgos_ACCESS_2020}
A.-A.~A. Boulogeorgos and A.~Alexiou, ``Performance analysis of reconfigurable
  intelligent surface-assisted wireless systems and comparison with relaying,''
  \emph{IEEE Access}, vol.~8, pp. 94\,463--94\,483, Jun. 2020.

\bibitem{Gan_LCOMM_2021}
X.~Gan, C.~Zhong, Y.~Zhu, and Z.~Zhong, ``User selection in reconfigurable
  intelligent surface assisted communication systems,'' \emph{IEEE Commun.
  Lett.}, 2021, {DOI: 10.1109/LCOMM.2020.3048782}.

\bibitem{Bjornson_LWC_2020_b}
E.~Bjornson and L.~Sanguinetti, ``Rayleigh fading modeling and channel
  hardening for reconfigurable intelligent surfaces,'' \emph{IEEE Wireless
  Commun. Lett.}, 2020, {DOI:} 10.1109/LWC.2020.3046107.

\bibitem{VanChienLWC2021}
T.~{Van Chien}, A.~K. Papazafeiropoulos, L.~T. Tu, R.~Chopra, S.~Chatzinotas,
  and B.~Ottersten, ``{Outage Probability Analysis of IRS-Assisted Systems
  Under Spatially Correlated Channels},'' \emph{IEEE Wireless Communications
  Letters}, Feb. 2021.

\bibitem{Ibrahim_TVT_2021}
H.~Ibrahim, H.~Tabassum, and U.~T.~Nguyen, ``Exact coverage analysis of
  intelligent reflecting surfaces with nakagami-m channels,'' \emph{IEEE Trans.
  Veh. Technol.}, vol.~70, no.~1, pp. 1072--1076, Jan. 2021.

\bibitem{Cui_TVT_2021}
Z.~Cui, K.~Guan, J.~Zhang, and Z.~Zhong, ``{SNR} coverage probability analysis
  of {RIS}-aided communication systems,'' \emph{IEEE Trans. Veh. Technol.},
  vol.~70, no.~4, pp. 3914--3919, Apr. 2021.

\bibitem{Jung_TWC_2020}
M.~Jung, W.~Saad, Y.~Jang, G.~Kong, and S.~Choi, ``Performance analysis of
  large intelligent surfaces {(LISs)}: Asymptotic data rate and channel
  hardening effects,'' \emph{{IEEE} Trans. Wireless Commun.}, vol.~19, no.~3,
  pp. 2052--2065, Mar. 2020.

\bibitem{Yang_WCL_2020}
L.~Yang, Y.~Yang, D.~B. da~Costa, and I.~Trigui, ``Outage probability and
  capacity scaling law of multiple {RIS}-aided networks,'' \emph{IEEE Wireless
  Commun. Lett.}, vol.~10, no.~2, pp. 256--260, Feb. 2021.

\bibitem{Fang_arXiv_2020}
\BIBentryALTinterwordspacing
Y.~Fang, S.~Atapattu, H.~Inaltekin, and J.~Evans, ``Optimum reconfigurable
  intelligent surface selection for indoor and outdoor communications,'' Dec.
  2020. [Online]. Available: \url{http://arxiv.org/abs/2012.11793}
\BIBentrySTDinterwordspacing

\bibitem{Mei_LWC_2020}
W.~Mei and R.~Zhang, ``Cooperative beam routing for multi-{IRS} aided
  communication,'' \emph{IEEE Wireless Commun. Lett.}, vol.~10, no.~2, pp.
  426--430, Feb. 2021.

\bibitem{Zhang_arXiv_2020}
\BIBentryALTinterwordspacing
S.~Zhang and R.~Zhang, ``Intelligent reflecting surface aided multiple access:
  Capacity region and deployment strategy,'' Feb. 2020. [Online]. Available:
  \url{http://arxiv.org/abs/2002.07091}
\BIBentrySTDinterwordspacing

\bibitem{Yildirim_TCOM_2020}
I.~Yildirim, A.~Uyrus, and E.~Basar, ``Modeling and analysis of reconfigurable
  intelligent surfaces for indoor and outdoor applications in future wireless
  networks,'' \emph{IEEE Trans. Commun.}, 2020, {DOI}:
  10.1109/TCOMM.2020.3035391.

\bibitem{Lyu_LWC_2020}
J.~Lyu and R.~Zhang, ``Spatial throughput characterization for intelligent
  reflecting surface aided multiuser system,'' \emph{IEEE Wireless
  Communications Letters}, vol.~9, no.~6, pp. 834--838, Jun. 2020.

\bibitem{Lyu_TWC_2021}
------, ``Hybrid active/passive wireless network aided by intelligent
  reflecting surface: System modeling and performance analysis,'' \emph{IEEE
  Trans. Wirel. Commun.}, 2021, {DOI:}10.1109/TWC.2021.3081447.

\bibitem{Gradshteyn2007}
I.~S. Gradshteyn and I.~M. Ryzhik, \emph{Tables of Integrals, Series, and
  Products}, 7th~ed.\hskip 1em plus 0.5em minus 0.4em\relax New York, NY, USA:
  Academic Press, 2007.

\bibitem{Abramowitz1965}
M.~Abramowitz and I.~A. Stegun, \emph{Handbook of Mathematical Functions: With
  Formulas, Graphs, and Mathematical Tables}.\hskip 1em plus 0.5em minus
  0.4em\relax New York, NY, USA: Dover, 1965.

\bibitem{Prudnikov1999}
A.~Prudnikov, Y.~Brychkov, and O.~Marichev, \emph{Integrals and Series, Volume
  3: More Special Functions}.\hskip 1em plus 0.5em minus 0.4em\relax Boca
  Raton, FL, USA: CRC Press, 1999.

\bibitem{Srivastava1985}
H.~M. Srivastava and P.~W. Karlsson, \emph{Multiple Gaussian Hypergeometric
  Series}.\hskip 1em plus 0.5em minus 0.4em\relax Hoboken, NJ, USA: Wiley,
  1985.

\bibitem{Bjornson_WCL_2020}
E.~Bjornson, O.~Ozdogan, and E.~G. Larsson, ``Intelligent reflecting surface
  versus decode-and-forward: How large surfaces are needed to beat relaying?''
  \emph{IEEE Wireless Commun. Lett.}, vol.~9, no.~2, pp. 244--248, Feb. 2020.

\bibitem{Zheng_LWC_2020}
B.~Zheng and R.~Zhang, ``Intelligent reflecting surface-enhanced {OFDM}:
  Channel estimation and reflection optimization,'' \emph{IEEE Wireless
  Communications Letters}, vol.~9, no.~4, pp. 518--522, 2020.

\bibitem{AlwazaniGLOBECOM2020}
H.~Alwazani, Q.-U.-A. Nadeem, and A.~Chaaban, ``Channel estimation for
  distributed intelligent reflecting surfaces assisted multi-user {MISO}
  systems,'' in \emph{2020 IEEE Globecom Workshops}.\hskip 1em plus 0.5em minus
  0.4em\relax IEEE, Dec. 2020, pp. 1--6.

\bibitem{Huang_TWC_2019}
C.~Huang, A.~Zappone, G.~C. Alexandropoulos, M.~Debbah, and C.~Yuen,
  ``Reconfigurable intelligent surfaces for energy efficiency in wireless
  communication,'' \emph{IEEE Trans. Wirel. Commun.}, vol.~18, no.~8, pp.
  4157--4170, Aug. 2019.

\bibitem{Yang_TVT_2020}
L.~Yang, F.~Meng, J.~Zhang, M.~O. Hasna, and M.~D. Renzo, ``On the performance
  of {RIS}-assisted dual-hop {UAV} communication systems,'' \emph{IEEE Trans.
  Veh. Technol.}, vol.~69, no.~9, pp. 10\,385--10\,390, Sep. 2020.

\bibitem{Matlab2021a}
\emph{{MATLAB version R2021a}}, The Mathworks, Inc., Natick, MA, USA, 2021.

\bibitem{Bowman1998}
K.~O. Bowman and L.~R. Shenton, \emph{Estimator: Method of Moments}.\hskip 1em
  plus 0.5em minus 0.4em\relax Hoboken, NJ, USA: Wiley, 1998.

\bibitem{Peebles2000}
P.~Z. Peebles, \emph{Probability, Random Variables and Random Signal
  Principles}, 4th~ed.\hskip 1em plus 0.5em minus 0.4em\relax New York, NY,
  USA: McGraw-Hill Science, 2000.

\bibitem{Bithas_TCOM_2020}
P.~S. Bithas, V.~Nikolaidis, A.~G. Kanatas, and G.~K. Karagiannidis,
  ``{UAV}-to-ground communications: {C}hannel modeling and {UAV} selection,''
  \emph{IEEE Trans. Commun.}, vol.~68, no.~8, pp. 5135--5144, Aug. 2020.

\bibitem{Stacy1962}
E.~W. Stacy, ``A generalization of the gamma distribution,'' \emph{The Annals
  of Mathematical Statistics}, vol.~33, no.~3, pp. 1187--1192, Sep. 1962.

\bibitem{Fenton_TCOM_1960}
L.~Fenton, ``The sum of log-normal probability distributions in scatter
  transmission systems,'' \emph{IEEE Trans. Commun.}, vol.~8, no.~1, pp.
  57--67, 1960.

\bibitem{Laourine_LCOMM_2007}
A.~Laourine, A.~St{\'{e}}phenne, and S.~Affes, ``Estimating the ergodic
  capacity of {Log-Normal} channels,'' \emph{IEEE Commun. Lett.}, vol.~11,
  no.~7, pp. 568--570, Jul. 2007.

\bibitem{3GPP}
``Evolved universal terrestrial radio access {(E-UTRA)}; further advancements
  for e-utra physical layer aspects (release 9),'' \emph{Document
  3GPP-TR-36.814 V9.0.0}, Mar. 2010.

\bibitem{AlAhmadiTWC2010}
S.~Al-Ahmadi and H.~Yanikomeroglu, ``On the approximation of the
  generalized-{K} distribution by a {Gamma} distribution for modeling composite
  fading channels,'' \emph{IEEE Trans. Wirel. Commun.}, vol.~9, no.~2, pp.
  706--713, Feb. 2010.

\bibitem{Kosti2005}
I.~Kosti, ``Analytical approach to performance analysis for channel subject to
  shadowing and fading,'' \emph{IEE Proceedings - Communications}, vol. 152,
  no.~6, p. 821, 2005.

\bibitem{Peppas_IET_2011}
K.~Peppas, ``Accurate closed-form approximations to generalised-{K} sum
  distributions and applications in the performance analysis of equal-gain
  combining receivers,'' \emph{IET Commun.}, vol.~5, no.~7, pp. 982--989, May
  2011.

\bibitem{KongACCESS2021}
L.~Kong, Y.~Ai, S.~Chatzinotas, and B.~Ottersten, ``{Effective Rate Evaluation
  of {RIS}-Assisted Communications Using the Sums of Cascaded $\alpha$-$\mu$
  Random Variates},'' \emph{IEEE Access}, vol.~9, pp. 5832--5844, 2021.

\bibitem{Costa_CL_2011}
D.~B. {da Costa}, H.~Ding, and J.~Ge, ``Interference-limited relaying
  transmissions in dual-pop cooperative networks over {Nakagami-m} fading,''
  \emph{IEEE Commun. Lett.}, vol.~15, no.~5, pp. 503--505, May 2011.

\bibitem{Abramowitz1972}
M.~Abramowitz and I.~A. Stegun, \emph{Handbook of Mathematical Functions with
  Formulas, Graphs, and Mathematical Tables}.\hskip 1em plus 0.5em minus
  0.4em\relax New York, NY, USA: Dover Books, 1972.

\end{thebibliography}
\balance
\end{document}